\documentclass[cup6b]{cupbook}

\title[ XXIII Canary Islands Winter School of Astrophysics]{Secular evolution of galaxies}
\author{J. Falc\'on-Barroso and J.~H.~Knapen} 
\date{20th March 2012}

\usepackage{amsmath}
\usepackage{amssymb}
\usepackage{graphicx}

\begin{document}

\maketitle


\def\thechapter{}

\author[Daniela Calzetti]{Daniela Calzetti\\  
        Department of Astronomy, University of Massachusetts,\\
        710 North Pleasant Street, Amherst, MA 01003, USA,\\
        calzetti@astro.umass.edu}

\chapter{Star formation rate indicators}

\abstract
{What else can be said about star formation rate indicators that has not been 
said already many times over? The `coming of age' of large ground-based surveys
and the unprecedented sensitivity, angular resolution and/or field-of-view of
infrared and ultraviolet space missions have provided extensive, homogeneous
data on both nearby and distant galaxies, which have been used to further our
understanding of the strengths  and pitfalls of many common star formation rate
indicators. The synergy between these surveys has also enabled the calibration
of indicators for use on scales that are comparable to those of star-forming
regions, thus much smaller than an entire galaxy. These are being used to
investigate star formation processes at the sub-galactic scale. I review 
progress in the field over the past decade or so.}
\newline\newline

\def\thechapter{1}

%
%

\section{Introductory remarks}
My goal for this chapter, based on a series of lectures at the XXIII Canary
Islands Winter School of Astrophysics, is to present  current understanding and
calibrations of star formation rate (SFR) indicators, both on global,
galaxy-wide scales, and on local, sub-galactic scales. SFRs are, together with
masses, the most important parameters that define galaxies and their evolution
across cosmic times. Although SFR calibrations have existed, with various levels
of accuracy, for many years and sometimes decades, the past eight to ten years
have brought forth major progress, through cohesive, multi-wavelength surveys
of nearby and distant galaxies. These surveys have exploited the sensitivity,
angular resolution and/or large field of view of  space telescopes (e.g., the
\textit{Spitzer Space Telescope}, the \textit{Galaxy Evolution Explorer}
[\textit{GALEX}], the \textit{Hubble Space Telescope} [\textit{HST}], and the
\textit{Herschel Space Telescope}) and leveraged the multi-band coverage
supplied by ground-based, all-sky surveys (e.g., the Sloan Digital Sky
Survey), in order to push the definition of SFR indicators into new regimes,
both in terms of wavelength coverage and spatial scales. I review here recent
progress in this area, but also highlight where challenges, sometimes unexpected
ones, have arisen. This chapter is structured to provide also a quick reference
for the relevant literature on SFR calibrations. It loosely follows the
structure of the lectures I presented at the Winter School. 

%
%

\section{Calibration of star formation rate indicators across the wavelength spectrum}
\label{calib}
Calibrations of SFR indicators have been presented in the literature for almost
30 years, derived across the full electromagnetic spectrum, from the X-ray,
through the ultraviolet (UV), via the optical and infrared (IR), all the way to
the radio, and using both  continuum and line emission (see review by Kennicutt
1998, and, e.g., Donas \& Deharveng 1984;  Yun {\it et al.} 2001; Kewley
{\it et al.} 2002, 2004; Ranalli {\it et al.} 2003; Bell 2003; Calzetti
{\it et al.} 2005, 2007, 2010; Schmitt {\it et al.} 2006; Alonso-Herrero {\it
et al.} 2006; Moustakas {\it et al.}  2006; Salim {\it et al.} 2007;
Persic \& Rephaeli 2007; Rosa-Gonz\'alez {\it et al.} 2007; Kennicutt {\it et
al.} 2007, 2009; Rieke {\it et al.} 2009; Lawton {\it et al.} 2010; Boquien {\it
et al.} 2010; Verley {\it et al.} 2010; Li {\it et al.} 2010; Treyer {\it et
al.} 2010; Murphy {\it et al.} 2011; Hao {\it et al.} 2011).  The most recent
review on the subject is by Kennicutt \& Evans (2012).

Recent findings that most of the star formation at redshift $z\sim$1--3  was
enshrouded in dust  (Le Floch {\it et al.} 2005; Magnelli {\it et al.} 2009;
Elbaz {\it et al.} 2011; Murphy {\it et al.} 2011b; Reddy {\it et al.} 2012)
have renewed interest in IR SFR indicators, particularly in the monochromatic
(single-bands) ones, which can be in principle as straightforward to use as
those already available at  UV and optical wavelengths. This interest has been
aided by the advent of high-angular resolution, high-sensitivity IR space
telescopes (\textit{Spitzer}, \textit{Herschel}), that have enabled the
calibration of monochromatic SFR indicators in nearby galaxies. The IR
investigations complement  efforts at UV and optical wavelengths to chart the
SFR evolution of galaxies from redshift $\sim$7--10 to the present  (e.g.,
Giavalisco {\it et al.} 2004; Bouwens {\it et al.} 2009, 2010). The UV and
optical may be the preferred SFR indicators at very high redshift, when galaxies
contained little dust (e.g., Wilkins {\it et al.} 2011; Walter {\it et al.}
2012). The calibration of SFR indicators  remains, however, a central issue for
studies of distant galaxies (e.g., Reddy {\it et al.} 2010; Lee {\it et al.}
2010; Wuyts {\it et al.} 2012), since it can be affected by differences in star
formation histories, metal abundances, content and distribution of stellar
populations and dust between low and high redshift galaxies (Elbaz {\it et al.}
2011), and, possibly, by cosmic variations in the cluster mass function and
stellar initial mass function (IMF, Wilkins {\it et al.} 2008; Pflamm-Altenburg
{\it et al.} 2009).   

Throughout this chapter, I refer to two categories of SFR calibrations: (1)
`global', i.e.,  defined for {\em whole galaxies}, thus they are
luminosity-weighted averages across local variations in star formation history
and physical conditions within each galaxy; and (2) `local', i.e., defined for
measuring SFRs in regions {\em within galaxies}, on sub-galactic/sub-kpc scale
(e.g., Wu {\it et al.} 2005; Alonso-Herrero {\it et al.} 2006; Calzetti {\it et
al.} 2005, 2007, 2010; Zhu {\it et al.} 2008; Rieke {\it et al.} 2009; Kennicutt
{\it et al.} 2009; Lawton {\it et al.} 2010; Boquien {\it et al.} 2010, 2011;
Verley {\it et al.} 2010; Li {\it et al.} 2010; Treyer {\it et al.} 2010; Liu
{\it et al.} 2011; Hao {\it et al.} 2011; Murphy {\it et al.} 2011). While
global SFR calibrations have received most of the attention in the past, both
for objective limitations in the spatial resolution of the data and for their
broader applicability to distant galaxy populations, local SFR calibrations have
become increasingly prominent  in the literature as important tools to
investigate the physical processes of star formation. 

The definition of a local SFR can, however, be problematic if referring to too
small a region: for instance a single star cluster that formed almost
instantaneously 15\,Myr ago has a current SFR=0 (it is no longer forming stars),
although stars were clearly formed in the recent past. To avoid such extreme
situations, local SFRs are meant to refer to measurements performed on areas
that include multiple star-forming regions, so that star formation can be
considered constant over the relevant timescale for the SFR indicator used. For
all practical purposes, such regions tend to be a few hundred pc across or
larger.

In general, global calibrations are not necessarily applicable to local
conditions and vice versa. The fundamental reason is that while the stellar and
dust emission from entire galaxies can be treated, in first approximation, as if
the galaxy were  an isolated system, the same is not necessarily true for a
sub-galactic region. Stellar populations mix within galaxies on timescales that
are comparable to those of their UV light lifetime. The stellar IMF (i.e., the
distribution of stellar masses at birth) may or may not be fully sampled
locally. The star formation history may vary from region to region. Both young
and old stars can heat the dust in a galaxy, and the dust spectral energy
distribution (SED) and features provide little discrimination as to the source
of the heating. Because of all these reasons, local SFR indicators are by far
less settled than the global ones. 

In what follows, I will discriminate between global and local SFR indicators,
when appropriate. All calibrations are given for a Solar metallicity stellar
population, when models are used. 

\subsection{General characteristics}
\label{general}
Techniques for measuring the rate at which stars are being formed vary
enormously, also depending on whether the target system is resolved into
individual units (e.g., young stars) or not. In all cases, however,  the basic
goal is to identify emission that probes newly or recently formed stars, while
avoiding as much as possible contributions from evolved stellar populations.

The timescale over which `recent' is a valid word also varies between different
applications and among different systems, but free-fall times $\tau_{\rm ff}$
probably provide a reasonable ballpark scale. Most researchers would agree that
`recent' refers to timescales $\approx$10--100\,Myr when considering whole
galaxies, and $\approx$1--10\,Myr when considering regions or structures within
galaxies (e.g., giant molecular clouds, etc.).

The most common approach for measuring SFRs in resolved regions, such as regions
within the Milky Way, is to count individual objects or events (e.g.,
supernovae) that trace the recent star formation (Chomiuk \& Povich 2011). In
the molecular clouds within 0.5--1\,kpc of the solar system, this is accomplished
by counting young stellar objects (YSOs), i.e., protostars at different stages
of evolution, which, because they are still embedded in their natal clouds, are
optimally identified in the IR. The total number of YSOs is converted to a SFR
via:
\begin{equation}
\label{eq:calzetti_1}
{\rm SFR(YSO)} = N_{\rm YSO} \frac{\langle M \rangle}{\tau}
\end{equation}
where the mean YSO mass, $\langle M \rangle$ depends weakly on the adopted
stellar IMF (see Section~\ref{IMF}), and the lifetime of a YSO is, with some
uncertainty,  $\tau\sim$2\,Myr (Evans {\it et al.} 2009; Heidermann {\it et al.}
2010; Gutermuth {\it et al.} 2011). The SFR(YSO) is in units of
$M_{\odot}$\,yr$^{-1}$. 

In unresolved systems, SFR indicators are merely measures of luminosity, either
monochromatic or integrated over some wavelength range, with the goal of
targeting continuum or line emission that is sensitive to the short-lived
massive stars. The conversion from the luminosity of massive stars to a SFR
is performed under the assumption that: (1) the star formation has been
roughly constant over the timescale probed by the specific emission being used;
(2) the stellar IMF is known (or is a controllable parameter) so that the number
of massive stars can be extrapolated to the total number of high$+$low mass
stars formed; and (3) the stellar IMF is fully sampled, meaning that at least
one star is formed in the highest-mass bin, and all other mass bins are
populated accordingly with one or more stars (see discussion in
Section~\ref{IMF}).

SFR indicators in the UV/optical/near-IR range
($\sim$0.1-5\,$\mu$m)  probe the direct stellar light emerging from galaxies,
while SFR indicators in the mid/far-IR ($\sim$5--1000\,$\mu$m) probe the stellar
light reprocessed by dust. In addition to direct or indirect stellar emission,
the ionising photon rate, as traced by the gas ionised by massive stars, can be
used to define SFR indicators; photo-ionised gas usually dominates over
shock-ionised gas in galaxies or large structures within galaxies (e.g.,
Calzetti {\it et al.} 2004; Hong {\it et al.} 2011). Tracers include hydrogen
recombination lines, from the optical, through the near-IR, all the way to radio
wavelengths, forbidden metal lines, and, in the millimetre range, the free-free
(Bremsstrahlung) emission. The X-ray emission produced by high-mass X-ray
binaries, massive stars, and supernovae can also, in principle, be used to trace
SFRs. Finally, the synchrotron emission from galaxies can be calibrated as a SFR
indicator (Condon 1992), since cosmic rays are produced and accelerated in
supernova remnants, and core-collapse supernovae represent 70\% or more of the
total supernovae in star-forming galaxies (Bossier \& Prantzos 2009). 

The following five subsections describe in more detail a few of these SFR
indicators for unresolved systems. The emission contribution to the galaxy
luminosity from a potential active galactic nuclei (AGN) can be large, depending
on the galaxy type and the wavelength of the SFR indicator. I assume that this
potential contribution has been recognised and removed from the emission that is
being used as a SFR indicator.

\subsubsection{Indicators based on direct stellar light}
\label{SFRUV}

The youngest stellar populations emit the bulk of their energy in the restframe
UV ($<$0.3\,$\mu$m); in the absence of dust attenuation, this is the wavelength
range `par excellence' to investigate star formation in galaxies over timescales
of $\approx$100--300\,Myr, since both O and B stars are brighter in the UV than
at longer wavelengths. As a reference, the lifetime of an O6 star is
$\sim$6\,Myr, and that of a B8 star is $\sim$350\,Myr.  The luminosity ratio at
0.16\,$\mu$m of an O6 to a B8 star is $\sim$90, but, if the stellar population
follows a Kroupa (2001) IMF (see Section~\ref{IMF}), for every O6 star formed,
about 150 B8 stars are formed. Thus, at age zero, the UV emission from the
collective contribution of B8 stars is comparable to that of O6 stars. 

For a Kroupa stellar IMF, with constant star formation over 100\,Myr, the
non-ionising UV ($0.0912\mu{\rm m}<\lambda<0.3\mu{\rm m}$) stellar continuum can be
converted to a SFR:
\begin{equation}
\label{eq:calzetti_2}
{\rm SFR(UV)} = 3.0\cdot10^{-47} \lambda L(\lambda)
\end{equation}
with SFR(UV) in $M_{\odot}$\,yr$^{-1}$,  $\lambda$ in \AA, and
\textit{L}($\lambda$) in erg/s. The stellar SED used for this calibration is
from Starburst99, with solar metallicity (Leitherer {\it et al.} 1999). The
accuracy of the calibration constant is $\pm$15\%, which takes into account
small variations as a function of $\lambda$. 

For constant star formation over timescales longer than 100\,Myr, the calibration
constant only decreases by a few percent. However, for shorter timescales,
changes are more significant. For $\tau$=10\,Myr and 2\,Myr, the constant is about
42\% and a factor 3.45, respectively, higher than in
Equation~\ref{eq:calzetti_2} (Table~\ref{tab:calzetti_1}). This shows that if
star formation has been active in a region on a timescale shorter than about
100\,Myr, the cumulative UV emission of massive stars is still increasing in
luminosity, and the calibration of any SFR(UV) indicator has to take this fact
into account.

\begin{table}[ht]
\begin{center}
\caption{\bf Model-based luminosity-to-SFR calibrations\label{tab:calzetti_1}}
\begin{tabular}{l l l }
\hline\hline
Luminosity$^a$ &$C^b$&Assumptions$^c$\\
\hline
\textit{L}(UV)         & 3.0$\times 10^{-47} \lambda$ & 0.1--100\,$M_{\odot}$, $\tau\ge$100\,Myr  \\ 
\textit{L}(UV)         & 4.2$\times 10^{-47} \lambda$ & 0.1--30\,$M_{\odot}$, $\tau\ge$100\,Myr  \\ 
\textit{L}(UV)         & 4.3$\times 10^{-47} \lambda$ & 0.1--100\,$M_{\odot}$, $\tau$=10\,Myr  \\ 
\textit{L}(UV)         & 1.0$\times 10^{-46} \lambda$ & 0.1--100\,$M_{\odot}$, $\tau$=2\,Myr  \\ 
\textit{L}(TIR)        & 1.6$\times 10^{-44}$         & 0.1--100\,$M_{\odot}$, $\tau$=10\,Gyr  \\ 
\textit{L}(TIR)        & 2.8$\times 10^{-44}$         & 0.1--100\,$M_{\odot}$, $\tau$=100\,Myr  \\ 
\textit{L}(TIR)        & 4.1$\times 10^{-44}$         & 0.1--30\,$M_{\odot}$, $\tau$=100\,Myr  \\ 
\textit{L}(TIR)        & 3.7$\times 10^{-44}$         & 0.1--100\,$M_{\odot}$, $\tau$=10\,Myr  \\ 
\textit{L}(TIR)        & 8.3$\times 10^{-44}$         & 0.1--100\,$M_{\odot}$, $\tau$=2\,Myr  \\ 
\textit{L}(H$\alpha$)  & 5.5$\times 10^{-42}$         & 0.1--100\,$M_{\odot}$, $\tau\ge$6\,Myr, $T_{\rm e}$=10$^4$\,k, $n_{\rm e}$=100\,cm$^{-3}$  \\ 
\textit{L}(H$\alpha$)  & 3.1$\times 10^{-41}$         & 0.1--30\,$M_{\odot}$, $\tau\ge$10\,Myr, $T_{\rm e}$=10$^4$\,k, $n_{\rm e}$=100\,cm$^{-3}$  \\ 
\textit{L}(Br$\gamma$) & 5.7$\times 10^{-40}$         & 0.1--100\,$M_{\odot}$, $\tau\ge$6\,Myr, $T_{\rm e}$=10$^4$\,k, $n_{\rm e}$=100\,cm$^{-3}$  \\ 
\hline
\end{tabular}
\footnotetext[1]{Luminosity in erg~s$^{-1}$. Stellar and dust continuum
luminosities are given as $\nu L$($\nu$); total IR=TIR is assumed to be equal to
the stellar population bolometric luminosity.}
\footnotetext[2]{The constant \textit{C} appears in the calibration as:
SFR($\lambda$)$=C L$($\lambda$), where SFR is in units of $M_{\odot}$\,yr$^{-1}$.
The constant is derived from stellar population models, with constant star
formation and solar metallicity (Starburst99, Leitherer {\it et al.} 1999). For
SFR(UV), the numerical value is multiplied by the wavelength $\lambda$ in \AA.}
\footnotetext[3]{Assumptions for mass range of the stellar IMF, which we adopt
to have the expression derived by Kroupa (2001), see Section~\ref{IMF}, and for
the timescale $\tau$ over which star formation needs to remain constant, for the
calibration constant to be applicable. For nebular lines, the adopted values of
electron temperature and density are also listed.}
\end{center}
\end{table}

A more subtle, but not less important, effect is caused by the length of time
over which a stellar SED remains relatively bright in the UV. This is due to the
significant UV emission of mid-to-late B stars. For example, a constant star
formation event of 10\,Myr duration, which, at constant
SFR=1\,$M_{\odot}$\,yr$^{-1}$, accumulates 10$^7$\,$M_{\odot}$ in stars, has the
same UV luminosity and a similar UV SED over the range 0.13--0.25\,$\mu$m of a
50\,Myr old, 2.5$\times$10$^8$\,$M_{\odot}$ instantaneous burst of star formation.
In the absence of dust attenuation and if only observed in the UV, the two
populations would be attributed the same SFR(UV)=1\,$M_{\odot}$\,yr$^{-1}$. While
this number is correct for the first population, it would be incorrect, and
possibly misleading, for the second population (which has not been forming stars
since 50\,Myr). If dust attenuation is also present, the potential of
misclassifying an ageing population for an active star-forming one increases.

As the vast majority of galaxies contain at least some dust, the use of SFR(UV)
becomes complicated, since dust attenuation corrections are usually required,
and are uncertain. For the most part dust corrections only work on ensembles of
systems, rather than individual objects. In a show of Cosmic Conspiracy, the
most active and luminous systems are also richer in dust, implying that they
require more substantial corrections for the effects of dust attenuation (Wang
\& Heckman 1996; Calzetti 2001; Hopkins {\it et al.} 2001; Sullivan {\it et al.}
2001; Calzetti {\it et al.} 2007). As a reference number, a modest optical
attenuation $A_V$=0.9 produces a factor ten reduction in the UV continuum at
0.13\,$\mu$m, if the attenuation curve follows the recipe of Calzetti {\it et
al.} (2000). 

\subsubsection{Indicators based on dust-processed stellar light}
\label{SFRIR}

The IR luminosity of a system will depend not only on its dust content, but also
on the heating rate provided by the stars. To first order, the shape of the
thermal IR SED will depend on the starlight SED, in the sense that UV-luminous,
young stars will heat the dust to higher mean temperatures than old stellar
populations (e.g., Helou 1986).

Because of the properties of the Planck function, hotter dust in thermal
equilibrium has higher emissivity in the IR than cooler dust. Furthermore, the
cross-section of the dust grains for stellar light is higher in the UV than in
the optical, as inferred from the typical trend of interstellar extinction
curves. Thus, qualitatively, the dust heated by UV-luminous, young stellar
populations will produce an IR SED that is more luminous and peaked at shorter
wavelengths (observationally $\approx$60\,$\mu$m) than the dust heated by
UV-faint, old or low-mass stars (observationally with an IR SED peak at
$\approx$100--150\,$\mu$m). This is the foundation for using the IR emission
($\sim$5--1000\,$\mu$m) as a SFR indicator.

The thermal IR emission is, however, a `blunt tool' for measuring SFRs, in the
sense that there is not a one-to-one mapping between UV photons and IR photons,
and a monochromatic heating source will produce a modified Planck function for
the thermal equilibrium dust emission. Hence the use of `bolometric' IR measures
for SFRs, where the IR emission is integrated over the full wavelength range; in
practice, most of the emission is located in the wavelength range
$\sim$5--1000\,$\mu$m. The bolometric IR luminosity is often indicated with
\textit{L}(TIR) (where TIR is the total infrared emission), and a star formation
rate calibration for a stellar population undergoing constant star formation
over $\tau$=100\,Myr is:
\begin{equation}
\label{eq:calzetti_3}
{\rm SFR(TIR)}=2.8\cdot10^{-44} L({\rm TIR})
\end{equation}
with SFR(TIR) in $M_{\odot}$\,yr$^{-1}$, and \textit{L}(TIR) in erg/s. For this
calculation I have assumed that the Starburst99, Solar-metallicity stellar
bolometric emission is completely absorbed and re-emitted by dust, i.e.,
$L_{\rm star}$(bol)=\textit{L}(TIR). 

Not all the stellar emission in a galaxy is generally absorbed by dust. A
ballpark number is given by the cosmic background radiation (e.g., Dole {\it et
al.} 2006), which shows about half of the light emerging at UV-optical-near-IR
wavelengths and half at IR wavelengths. Thus, a simplified approach would be to
assume that in a typical galaxy only about half of its stellar light is absorbed
by dust. This fraction is, however, strongly dependent on the dust content and
distribution within the galaxy itself. The application of the SFR(TIR)
calibration derived in this section to actual galaxies, which is based on models
and the assumption that {\em all} of the stellar emission is absorbed by dust
and re-emitted in the IR,  will therefore result in a lower limit to the true
SFR. 

The main reason for giving a theoretical expression for SFR(TIR) is to show how
dependent the calibration is on assumptions on the stellar population's
characteristics. If $\tau$=10\,Myr and 2\,Myr, the calibration constant has
$\tau$-dependent variations that are not too dissimilar from those of SFR(UV)
(Table~\ref{tab:calzetti_1}). However, unlike SFR(UV), the calibration constant
of SFR(TIR) keeps changing for star formation timescales longer than 100\,Myr,
and for $\tau$=10\,Gyr it is about 57\% of the 100\,Myr calibration constant. The
difference relative to the SFR(UV) case is due to the accumulation over time of
long-lived, low-mass stars in the stellar SED. These contribute to the TIR
emission, but not to the UV one. The heating of dust by multiple-age stellar
populations has the additional effect of producing a thermal equilibrium IR SED
that is significantly broader than that produced by a single-temperature
modified blackbody function. This has been modelled in the past with at least
two approaches: (1) two or more dust components with different temperatures, or
(2) one single-temperature dust component with a small absolute value of the
dust emissivity index. Physically-motivated models are now available (Draine \&
Li 2007), which describe the dust emission from galaxies with large accuracy
(Draine {\it et al.} 2007; Aniano {\it et al.} 2012).

Progress over the past $\sim$10--20 years in dissecting the various dust
components that contribute to the IR SED has helped refining the original simple
picture (of which a summary can be found in, e.g., Draine 2003, 2009). Since
this chapter is about SFR indicators and not dust properties, I will only
summarise the salient traits that connect dust characteristics to wavelength
regions in the TIR emission.

The short-wavelength mid-IR range ($\sim$3--20\,$\mu$m) dust emission arises from
a combination of broad emission features, generated by the bending and
stretching modes of polycyclic aromatic hydrocarbons (PAHs), and continuum. The
latter is due to emission from both single-photon, stochastically-heated small
dust grains and thermally emitting hot (T$>$150\,K) dust: which of these two
components dominates depends on the nature of the heating sources, although
single-photon heating is predominant in the general interstellar medium of the
Milky Way. 

The long-wavelength mid-IR range ($\sim$20--60\,$\mu$m) is emission continuum
dominated by hot/warm (T$\ge$50\,K) dust in thermal equilibrium and single-photon
heated small-grain dust. This is the region where, in most galaxies, the dust
emission transitions from being dominated by emission from stochastically heated
grains to being dominated by large grains in thermal equilibrium. 

Finally, the far-IR range ($\gtrsim$60\,$\mu$m) is mainly due to thermal emission
from large grains. The mean temperature of the dust decreases for longer
wavelengths (termed `cool' or `cold' dust, depending on the author), although
typical temperatures are about 15--20\,K or above. 

Both massive, short-lived stars and low-mass, long-lived stars can heat the dust
contributing to each of the spectral regions identified above. However,
UV-bright stars will likely heat the surrounding dust to relatively high
effective temperatures. The $\sim$20--60\,$\mu$m  IR wavelength region, where the
emission transitions from stochastic heating to thermal heating, has thus been
targeted as a promising region for defining monochromatic (single-band) SFR
indicators. The advantage of such indicators is the ease of use: instead of
obtaining multi-point measurements along the IR SED and/or perform uncertain
extrapolations, monochromatic IR SFR indicators only require a single wavelength
measurement. 

Owing to the uncertainty of assigning a given waveband to a specific dust
emission component, monochromatic IR SFRs have been calibrated across a wide
range of IR wavelengths, including the \textit{IR Space Observatory}
(\textit{ISO}) 7 and 15\,$\mu$m bands, the \textit{Spitzer Space Telescope} 8,
24, 70\,$\mu$m bands, and, currently, the \textit{Herschel Space Telescope}
70\,$\mu$m and longer wavelength bands. The range of   angular resolutions
offered by each facility has been and is enabling the calibration of both global
and local SFR indicators. 

Monochromatic SFR indicators shortward of 15--20\,$\mu$m require care of use:
stochastically-heated dust can trace both young and evolved stellar populations
(e.g., Boselli {\it et al.} 2004; Calzetti {\it et al.} 2007; Bendo {\it et al.}
2008; Crocker {\it et al.} 2012). The PAHs may be better tracers of B stars than
current SFR (Peeters {\it et al.} 2004), and the  emission features show strong
dependence on the metal abundance of the system (e.g., Madden {\it et al.} 2000,
2006;  Engelbracht {\it et al.} 2005, 2008; Draine {\it et al.} 2007; Smith {\it
et al.} 2007; Galliano {\it et al.} 2008; Gordon {\it et al.} 2008;
Mu\~noz-Mateos {\it et al.} 2009; Marble  {\it et al.} 2010). Only about 50\%
of the emission at 8\,$\mu$m from a galaxy is dust-heated by stellar populations
10\,Myr or younger, and about 2/3 by stellar populations 100\,Myr or younger 
(Crocker {\it et al.} 2012). Thus, a significant fraction of the 8\,$\mu$m
emission is unrelated to current star formation. While this is likely to affect
mainly studies of sub-galactic regions or structures, some effect may be
expected on the global SFR indicators. Various calibration efforts have usually
recovered a roughly linear or slightly sub-linear relation between the global
8\,$\mu$m luminosity, providing reference SFR indicators for metal-rich, 
star-forming galaxies. However, the peak-to-peak scatter tends to be large, a
factor of about three (Treyer {\it et al.} 2010), with  larger deviations 
observed for compact starbursts (Elbaz {\it et al.} 2011). 

Longward of around 70\,$\mu$m, the contribution to the IR SED of thermal dust at
increasingly lower temperature, and therefore heated by stars that are low-mass
and long-lived, becomes more and more prominent, thus compromising the ability
of the IR emission to trace exclusively or almost exclusively recent star
formation. 

For the above two reasons, I offer here only two {\em empirical} monochromatic
IR SFR calibrations: in the 24\,$\mu$m and 70\,$\mu$m restframe bands. I
distinguish between local and global SFR indicators, since, as discussed earlier
in this section, the bolometric luminosity of a stellar population undergoing
constant star formation increases with time. Therefore, to the extent that the
IR emission traces the bolometric emission of the stellar population, the
calibration constant will be different for a global, galaxy-wide SFR indicator
and a local SFR indicator, since the former includes the
Hubble-time-integrated stellar population of a galaxy, while the latter is
generally derived from regions that are dominated by stellar populations with
short star formation timescales (H{\sc ii} regions, large star-forming complexes,
etc.). 

At 24\,$\mu$m, the local (spatial scale $\sim$500\,pc) calibration offered by
Calzetti {\it et al.} (2007) is:
\begin{equation}
\label{eq:calzetti_4}
{\rm SFR(24)_{local}}= 1.31\cdot 10^{-38} L(24)^{0.885}\hfil~~~~(1\cdot 10^{40}\lesssim L(24)\lesssim 3\cdot 10^{44})
\end{equation}
with SFR(24) in $M_{\odot}$\,yr$^{-1}$, and $L(24)=\nu L$($\nu$) in
erg~s$^{-1}$. The uncertainty is 0.02 in the exponent, and 15\% in the
calibration constant. The non-linear correlation between \textit{L}(24) and SFR
is a common characteristic of this tracer at the local scale (Alonso-Herrero
{\it et al.} 2006; P\'erez-Gonz\'alez {\it et al.} 2006; Calzetti {\it et al.}
2007; Rela\~no {\it et al.} 2007; Murphy {\it et al.} 2011), and may be a
manifestation of the increasing transparency of regions for decreasing
\textit{L}(24) luminosity, of the increasing  mean dust temperature for
increasing \textit{L}(24) luminosity, or a combination of the two. 

At the galaxy-wide scale, both linear and non-linear correlations between SFR
and \textit{L}(24) have been derived (Wu {\it et al.} 2005; Zhu {\it et al.}
2008; Kennicutt {\it et al.} 2009; Rieke {\it et al.} 2009), perhaps owing to
differences in sample selections and in the reference SFR indicators used to
calibrate SFR(24), and sufficient scatter in the data that both linear and
non-linear fits can be accommodated (Wu {\it et al.} 2005; Zhu {\it et al.}
2008). The linear  calibration of Rieke {\it et al.} (2009), reported using the
same IMF as all the other indicators in this presentation, is:
\begin{eqnarray}
\label{eq:calzetti_5}
{\rm SFR(24)_{global}}&=& 2.04\cdot 10^{-43} L(24)~~~~~~~~~~~~(4\cdot10^{42}\le L(24)\le 5\cdot 10^{43})\nonumber \\
                      &=& 2.04\cdot 10^{-43} L(24)  \nonumber \\ 
                      &\times& [2.03\cdot 10^{-44}L(24)]^{0.048}~~~~~(L(24)>5\cdot10^{43})
\end{eqnarray}
where a small correction for self-absorption is included at the high
luminosities. The linear calibrations in the literature tend to be within 30\%
of each other, suggesting a general agreement. 

\begin{figure}  
\includegraphics[width=0.9\linewidth]{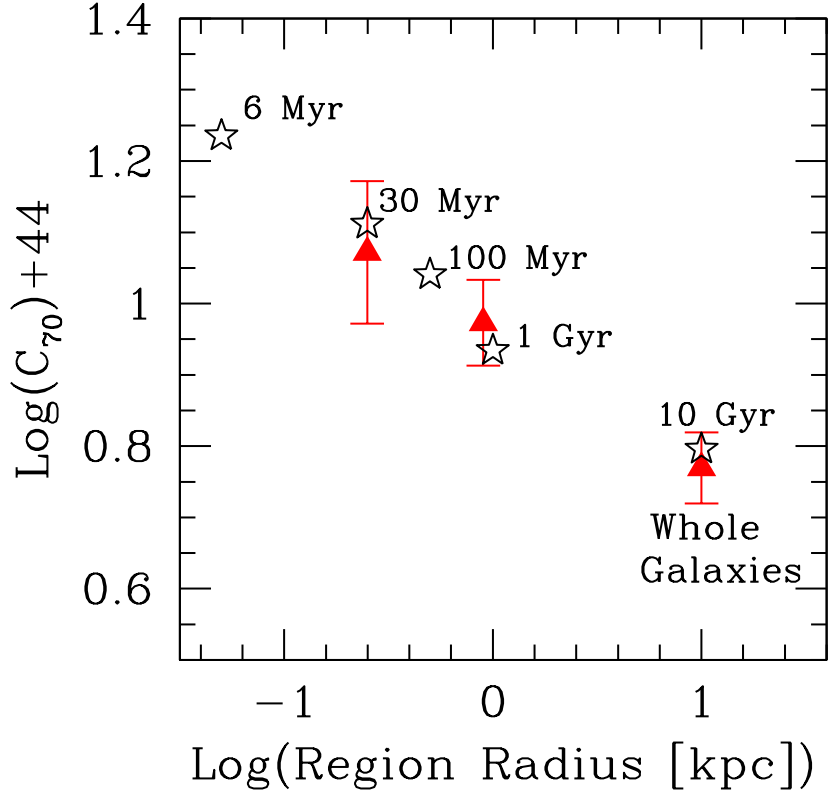}
\caption{The calibration constant, $C_{70}$,  between SFR and the 70\,$\mu$m
luminosity, expressed as SFR(70)=$C_{70}$\ \textit{L}(70), as a function of the
physical size of the regions used to derive the calibration. The filled red
triangles are observed values from Li {\it et al.} (2010, 2012) and Calzetti
{\it et al.} (2010), using both \textit{Spitzer} and \textit{Herschel} data. The
black stars are from stellar population synthesis models, for  constant star
formation and a Kroupa IMF, in the stellar mass range 0.1--100\,$M_{\odot}$; the
mean age of the population that best approximates the observed $C_{70}$ values
is shown. The scaling between bolometric light and 70\,$\mu$m emission is
discussed in Calzetti {\it et al.} (2010). The association between star
formation timescale and region size is based on a region crossing time with a
1--3\,km~s$^{-1}$ speed.}
\label{fig01}
\end{figure}

At 70\,$\mu$m, the local calibration over $\sim$1\,kpc scales derived by Li {\it
et al.} (2010) using over 500 star-forming regions is:
\begin{equation}
\label{eq:calzetti_6}
{\rm SFR(70)_{local}}= 9.4\cdot10^{-44} L(70)~~~~~~~~~(5\times10^{40}\lesssim L(70)\lesssim 5\times 10^{43})
\end{equation}
with SFR(70) in $M_{\odot}$\,yr$^{-1}$, and \textit{L}(70)=$\nu L$($\nu$) in erg~s$^{-1}$.
The formal uncertainty on the calibration constant is about 2\%, although the
scatter in the datapoints is about 35\%. The global calibration provided by
Calzetti {\it et al.} (2010) is:
\begin{equation}
\label{eq:calzetti_7}
{\rm SFR(70)_{global}}= 5.9\cdot10^{-44} L(70)~~~~~~~~~(L(70)\gtrsim 1.4\times 10^{42})
\end{equation}
with a scatter of the datapoints of about 60\%. The SFR(70) calibration constant
thus increases when going from whole galaxies to 1\,kpc regions, i.e., for
decreasing region sizes. Li {\it et al.} (2012) obtain a tantalising result: the
calibration constant for SFR(70) becomes even larger for regions smaller than
$\sim$1\,kpc. These constants can be interpreted in terms of star formation
timescale within each region size (Fig.~\ref{fig01}), for simple assumptions
on the star formation history,  the fraction of stellar light re-emitted by
dust in the IR, and  the fraction of IR emission contained in the 70\,$\mu$m band
(Draine \& Li 2007). 

Gas-rich, but metal-poor, galaxies offer little opacity to stellar emission (the
same is true for metal-rich, but gas-poor galaxies, e.g., ellipticals, but for
these galaxies few or no stars form). Metal content correlates with galaxy
luminosity (Tremonti {\it et al.} 2004, and references therein), and faint
galaxies are faint IR emitters. In this case, SFR(IR) becomes a highly uncertain
tool. SFR tracers that mix tracers of both dust-obscured and dust-unobscured
star formation have recently been calibrated, and will be presented in
Section~\ref{SFRMIX}.

\subsubsection{Indicators based on ionised gas emission}
\label{SFRION}

Young, massive stars produce copious amounts of ionising photons that ionise the
surrounding gas. Hydrogen recombination cascades produce line emission, 
including the well-known Balmer series lines of H$\alpha$ (0.6563\,$\mu$m) and
H$\beta$ (0.4861\,$\mu$m), which, by virtue of being strong and located in the
optical wavelength range, represent the most traditional SFR indicators
(Kennicutt 1998). 

Only stars more massive than $\sim$20\,$M_{\odot}$ produce a measurable ionising 
photon flux. In a stellar population formed through an instantaneous burst with 
a Kroupa IMF the ionising photon flux decreases by two orders of magnitude 
between 5\,Myr and 10\,Myr after the burst. 

The relation between the intensity of a hydrogen recombination line and the
ionising photon rate is dictated by quantum mechanics, for a nebula that is
optically thick to ionising photons (case B, Osterbrock \& Ferland 2006). Case B
is usually assumed for most astrophysical situations where SFRs are of interest.
Although typical interstellar hydrogen densities are sufficient to ensure that
most H{\sc ii} regions should be radiation-bound, inhomogeneities in the
interstellar medium cause some fraction of the ionising photons to leak out of
the regions (but usually not out of galaxies). More discussion on this issue is
given below. The relation between the luminosity of the H$\alpha$ emission line
and the ionising photon rate is given by Osterbrock \& Ferland (2006):
\begin{equation}
\label{eq:calzetti_8}
L({\rm H}\alpha) = \frac{\alpha_{{\rm H}\alpha}^{\rm eff}}{\alpha_{\rm B}} h \nu_{{\rm H}\alpha} Q({\rm H}^o) = 1.37\cdot10^{-12}\ Q({\rm H}^o)
\end{equation}
where \textit{L}(H$\alpha$) is in erg~s$^{-1}$, $\alpha_{{\rm H}\alpha}^{\rm
eff}$ is the effective recombination coefficient at H$\alpha$, $ \alpha_{\rm B}$
is the case B recombination coefficient, $Q({\rm H}^o)$ is the ionising photon
rate in units of s$^{-1}$, and the constant at the right-hand side of the
equation is the resulting coefficient for electron temperature $T_{\rm
e}$=10000\,K and density $n_{\rm e}$=100\,cm$^{-3}$. For a Kroupa IMF
(Section~\ref{IMF}), the relation between ionising photon rate and SFR is:
\begin{equation}
\label{eq:calzetti_9}
{\rm SFR}(Q({\rm H}^o)) = 7.4\cdot10^{-54} Q({\rm H}^o)
\end{equation}
with SFR(Q$({\rm H}^o)$) in $M_{\odot}$~yr$^{-1}$. Combining
Equations~\ref{eq:calzetti_8} and \ref{eq:calzetti_9}, we obtain the well-known
calibration:
\begin{equation}
\label{eq:calzetti_10}
{\rm SFR}({\rm H}\alpha) = 5.5\cdot10^{-42} L({\rm H}\alpha)
\end{equation}
again, with SFR(H$\alpha$) in $M_{\odot}$~yr$^{-1}$ and \textit{L}(H$\alpha$) in
erg~s$^{-1}$. The variation of the calibration constant is $\sim$15\% for
variations in electron temperature in the range $T_{\rm e}$=5000--20000\,K ,
and is $<$1\% for electron density variations in the range $n_{\rm
e}$=100--10$^6$\,cm$^{-3}$ (Osterbrock \& Ferland 2006). Star formation needs to
have remained constant over timescales $>$6\,Myr for the calibration constant to
be applicable (Table~\ref{tab:calzetti_1}), but there is no dependency on long
timescales, unlike SFR(UV) or SFR(TIR). 

All SFR indicators that use the ionisation of hydrogen to trace the formation of
massive stars are sensitive to the effects of dust. The most commonly treated
effect is that of dust attenuation of the line or continuum. As we will see in
Section~\ref{DUSTATTEN}, various techniques have been developed to try to remove
this effect; furthermore, dust attenuation decreases for increasing wavelength.
A far more difficult effect to treat is the direct absorption of Lyman continuum
photons by dust. In this case, the ionising photons are removed altogether from
the light beam and are no longer available to ionise hydrogen. Thus, no emission
from either recombination lines or free-free continuum emission will result. The
actual impact of Lyman continuum photon (Lyc) absorption by dust has been
notoriously difficult to establish from an empirical point of view, owing to the
absence of a `ground truth' (or reference) with which to compare measurements.
Models have to be involved, and these show that the level of Lyc absorption
depends on the assumption for the geometry of the nebulae (Dopita {\it et al.}
2003, and references therein). The parametrisation of Dopita {\it et al.}, where
the ratio of H$\beta$ line luminosity with and without Lyc absorption is given
as a function of the product of metal abundance and ionisation parameter, shows
that most normal disk galaxies fall into the regime of low Lyc absorption,
typically less than 15\%--20\%; however, Lyc absorption by dust can become
significant at large ionisation parameters and metallicities, such as those
typical of local luminous and ultra-luminous IR galaxies (LIRGs and ULIRGs,
galaxies with bolometric luminosity $>$~a~few~10$^{11}$\,$L_{\odot}$), and of
some high-density central regions of galaxies. 

A somewhat opposite effect is represented by leakage of ionising photons, i.e.,
case B recombination does not fully apply. Leakage of ionising photons from
galaxies is likely negligible, at the level of a few percent or less (e.g.,
Heckman {\it et al.} 2011), although the jury is still out in the case of
low-mass, low-density galaxies (Hunter {\it et al.} 2010; Pellegrini {\it et
al.} 2012). Star-forming regions within galaxies tend, on the other hand, to be
leaky, and lose about  25\%--40\% of their ionising photons (see recent work by
Pellegrini {\it et al.} 2012; Rela\~no {\it et al.} 2012; Crocker {\it et al.}
2012). Thus, the use of ionising photon tracers for local SFRs may be biased
downwards by about 1/3 of their true value because of this effect. This
correction is not included in Table~\ref{tab:calzetti_1}.

Adopting for the moment that Lyc absorption by dust and leakage are not issues,
emission lines still need to be corrected for the effects of dust attenuation.
As an example, a modest attenuation of $A_V$=1\,mag by foreground dust depresses
the H$\alpha$ luminosity by a factor $\sim$2. At longer wavelengths, Br$\gamma$
(2.16\,$\mu$m) is depressed by only 11\%, for the same $A_V$. Recent advances in
the linearity, stability, and field-of-view size of infrared detectors' are
making it possible to collect significant samples of galaxies observed in the IR
Hydrogen recombination lines. Table~\ref{tab:calzetti_1} shows a calibration for
Br$\gamma$, derived under the same assumptions as SFR(H$\alpha$). Calibrations
for other lines can be inferred from those of H$\alpha$ and Br$\gamma$ and the
emissivity ratios published in Osterbrock \& Ferland (2006). 

Recombination lines at wavelengths longer than the optical regime, while
offering the advantage of lower sensitivity to dust attenuation,  have the dual
disadvantage of being progressively fainter and more sensitive to the physical
conditions of the gas, especially the temperature; these are natural
consequences of transition probabilities and conditions of thermal equilibrium,
respectively. The luminosity of Br$\gamma$ is about 1/100th of that at
H$\alpha$, and it changes by about 35\% for $T_{\rm e}$ in the range
5000--20000\,K, and by $\sim$4\% for density in the range $n_{\rm
e}$=10$^2$--10$^6$\,cm$^{-3}$. For Br$\alpha$ (4.05\,$\mu$m), the variations are
58\% and 13\% for changes in $T_{\rm e}$ and $n_{\rm e}$, respectively. The
sensitivity of Br$\gamma$ and Br$\alpha$ to $T_{\rm e}$ is a factor 2.4 and 3.9
larger, respectively, than that of \textit{L}(H$\alpha$).\looseness-2 

New or greatly improved radio and millimetre facilities such as
\textit{ALMA} (the Atacama Large Millimetre/submillimetre Array) or
\textit{EVLA} (the Expanded Very Large Array) are opening the window for
exploring millimetre and/or radio recombination lines  as ways to measure SFRs
unimpeded by effects of dust attenuation, albeit using lines that are
intrinsically extremely weak. I will cumulatively refer as RRLs all
(sub)millimetre and radio recombination lines from hydrogen quantum levels
$n>$20. At high quantum numbers ($n>$80--200, depending on electron density),
i.e., at wavelengths of a few cm or longer, stimulated emission is no longer
negligible and adds extra parameters in the expression of the line luminosity
(Brown {\it et al.} 1978). Even within the regime where stimulated emission is
not a concern, the line luminosity is dependent on the electron temperature,
producing SFR(RRL)$\propto T_{\rm e}^{0.7}$ (Gordon \& Sorochenko 2009). This
translates into a variation in the line intensity of a factor 2.6 for $T_{\rm
e}$ in the range 5000--20000\,K. 

The traditional approach of measuring the electron temperature from the radio
line-to-continuum ratio relies on the assumption that the underlying continuum
is free-free emission. The true level of free-free emission from a galaxy, or
from a large star-forming region embedded in a galaxy, needs to be carefully
disentangled from both dust emission (dominant $<$~2--3\,mm) and
synchrotron emission (dominant $>$~1--3\,cm, depending on the source). Often,
multi-wavelength observations are used to accomplish this (e.g., Murphy {\it et
al.} 2011), and add an additional layer of complication to the use of SFR(RRL).
Exploratory work is still ongoing to test how efficiently RRLs can be detected
in external galaxies (e.g., Kepley {\it et al.} 2011), and what advantage they
can bring relative to more classical and efficient methods. If the high-density,
and heavily dust-obscured, regime turns out to be the main niche for these
tracers (Yun 2008), they will need to be carefully weighted against SFR(IR), in
light of the potentially heavy impact of the Lyc absorption by dust. 

The free-free emission from galaxies or regions itself is a SFR tracer, being
the product of the Coulomb interaction between free electrons and ions in
thermal equilibrium. The electron temperature dependence of this SFR tracer,
SFR(ff)$\propto T_{\rm e}^{-0.45}$ (Condon 1992), is shallower than that of the
SFR(RRL), implying less than a factor of two change for a factor of four
variation in $T_{\rm e}$. A calibration of SFR(ff) consistent with our IMF
choice is given in Murphy {\it et al.} (2011).  

SFR tracers that use forbidden metal line emission will not be discussed in this
review, as they suffer from the same limitations as the hydrogen recombination
lines, and have additional dependencies on the metal content and ionisation
conditions of a galaxy or region. A review is found in Kennicutt (1998) and a
recent calibration in Kennicutt {\it et al.} (2009). 

\subsubsection{Indicators based on mixed processes}
\label{SFRMIX}
\begin{figure}  
\includegraphics[width=0.98\linewidth]{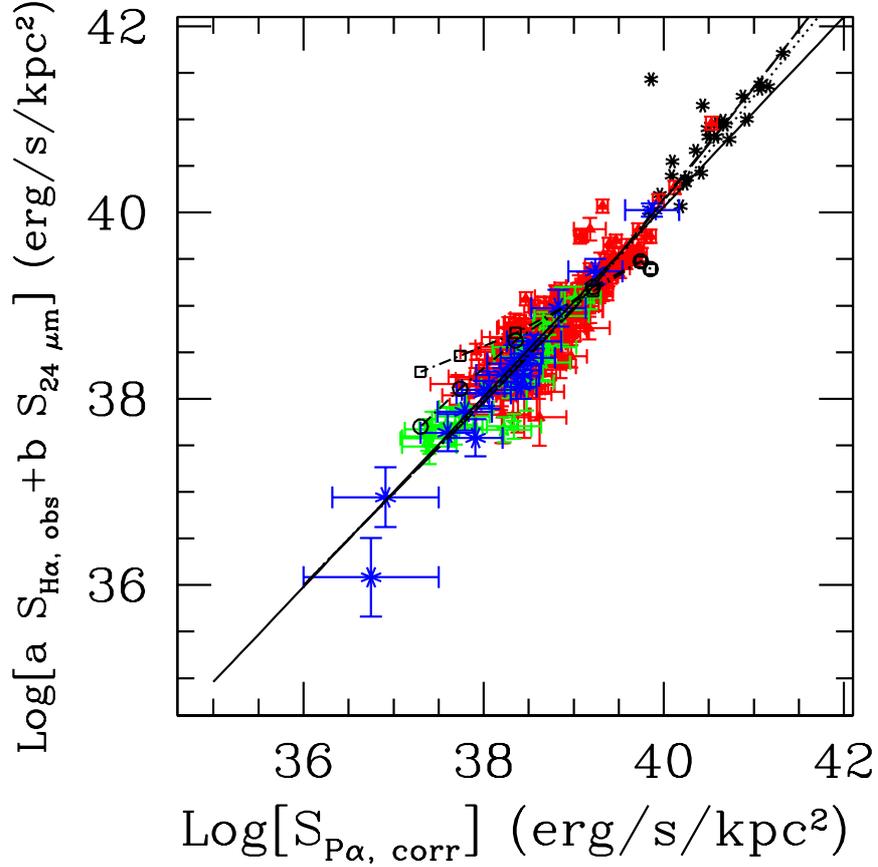}
\caption{An example of the calibration for a mixed SFR indicator, from Calzetti
{\it et al.} (2007). This specific example is for a {\it local} SFR indicator:
the data points include star-forming regions in nearby galaxies (red triangles,
green squares, blue crosses) and local LIRGs (black stars, from Alonso-Herrero
{\it  et al.} 2006). The horizontal axis is the luminosity/area of the
regions/galaxies in the hydrogen recombination line P$\alpha$ (1.8756\,$\mu$m),
the vertical axis is a linear combination of the luminosity/area of the observed
H$\alpha$ and 24\,$\mu$m luminosity. The star-forming regions include low (blue),
intermediate (green), and high (red) metallicity. All the points align basically
along a one-to-one relation (straight continuous line) suggesting that this
calibration is fairly independent of the metallicity (dust content) and
luminosity of the source. Other lines mark the position of models as described
in Calzetti {\it et al.} (2007).}
\label{fig02}
\end{figure}

The necessity to capture both dust-obscured and dust-unobscured star formation has
led to the formulation of SFR indicators that attempt to use the best qualities
of each indicator above. This advantage compensates for the slight disadvantage
of having to obtain two measures at, generally, two widely separated
wavelengths: one that measures the direct stellar light and one that measures
the dust-processed light. These mixed indicators have been calibrated
empirically over the past $\sim$5--7 years, using combinations of data from
space and the ground (e.g., Calzetti {\it et al.} 2005, 2007; Kennicutt {\it et
al.} 2007, 2009; Liu {\it et al.} 2011; Hao {\it et al.} 2011), and are usually
expressed as:
\begin{equation}
\label{eq:calzetti_11}
{\rm SFR}(\lambda_1,\lambda_2) = C(\lambda_1) [L(\lambda_1)_{\rm obs} + a_{\lambda_2,{\rm Type}} L(\lambda_2)_{\rm obs}]
\end{equation}
with SFR$(\lambda_1,\lambda_2)$ in $M_{\odot}$~yr$^{-1}$. $\lambda_1$ is usually
a wavelength probing either direct stellar light (e.g., the \textit{GALEX} FUV
at 0.153\,$\mu$m) or ionised gas tracers (e.g., H$\alpha$), and $\lambda_2$ is a
wavelength or range of wavelengths where dust emission dominates (e.g.,
24\,$\mu$m, 25\,$\mu$m, TIR, etc.). The constant $C(\lambda_1)$ is the calibration
for the direct stellar light probe, often derived from models (see
Table~\ref{tab:calzetti_1}). The luminosities $L(\lambda_1)_{\rm obs}$ and
$L(\lambda_2)_{\rm obs}$ are in units of erg~s$^{-1}$ and are the {\em observed}
luminosities (i.e., not corrected for effects of dust attenuation or other
effects). The proportionality constant a$_{\lambda_2, {\rm Type}}$ depends on
both the dust emission tracer used and whether the calibration is for local or
global use (Type=local, global). This latter characteristic is due to the
sensitivity of dust emission to heating from a wide range of stellar populations
(see Section~\ref{SFRIR}). An example of a mixed indicator calibration is given
in Fig.~\ref{fig02}.

\begin{table}[ht]
\begin{center}
\caption{\bf Mixed processes-based SFR calibrations$^a$\label{tab:calzetti_2}}
\begin{tabular}{l l l l l}
\hline\hline
$\lambda_1^b$ & $\lambda_2^b$ &$C_{\lambda_1}^c$& $a_{\lambda_2,{\rm Type}}^d$ & Type$^d$\\
\hline
FUV (0.153\,$\mu$m) & 24\,$\mu$m & 4.6$\times$10$^{-44}$ & 6.0    & local \\
FUV (0.153\,$\mu$m) & 25\,$\mu$m & 4.6$\times$10$^{-44}$ & 3.89   & global \\
FUV (0.153\,$\mu$m) & TIR       & 4.6$\times$10$^{-44}$ & 0.46   & global \\
H$\alpha$          & 24\,$\mu$m & 5.5$\times$10$^{-42}$ & 0.031  & local \\
H$\alpha$          & 25\,$\mu$m & 5.5$\times$10$^{-42}$ & 0.020  & global \\
H$\alpha$          & TIR       & 5.5$\times$10$^{-42}$ & 0.0024 & global \\
\hline
\end{tabular}
\footnotetext[1]{The calibrations are expressed as: SFR$(\lambda_1,\lambda_2) =
C(\lambda_1) [L(\lambda_1)_{\rm obs} + a_{\lambda_2,{\rm  Type}}
L(\lambda_2)_{\rm obs}]$, see text. The SFR is in units of
$M_{\odot}$~yr$^{-1}$. The luminosities $L(\lambda_1)_{\rm obs}$ and
$L(\lambda_2)_{\rm obs}$ are in units of erg~s$^{-1}$ and are the {\em observed}
luminosities.  Stellar or dust continuum luminosities are given as $\nu
L$($\nu$); TIR is the dust luminosity integrated in the range
$\sim$5--1000\,$\mu$m (e.g., Dale \& Helou 2002). The calibration
constants are from: Calzetti {\it et al.} 2005, 2007; Kennicutt {\it et al.}
2007, 2009; Liu {\it et al.} 2011; Hao {\it et al.} 2011.}
\footnotetext[2]{$\lambda_1$ is centred  in the \textit{GALEX} FUV at
0.153\,$\mu$m  or at H$\alpha$, and $\lambda_2$ is  24\,$\mu$m, 25\,$\mu$m, or TIR.
The difference in luminosity between 24\,$\mu$m and 25\,$\mu$m is around 2\%
(Kennicutt {\it et al.} 2009; Calzetti {\it et al.} 2010).} 
\footnotetext[3]{The constant $C(\lambda_1)$ is  the calibration for the direct
stellar light probe, derived from models or empirically.   In this Table, we
adopt model-derived values (Table~\ref{tab:calzetti_1}).} 
\footnotetext[4]{The constant $a_{\lambda_2,{\rm Type}}$ provides the fraction
of dust-processed light at $\lambda_2$ that needs to be added to the direct
stellar/gas probe at $\lambda_1$. `Type' refers to either a local calibration
(applicable to regions in galaxies $\lesssim$0.5--1\,kpc) or a global calibration
(whole galaxies).}
\end{center}
\end{table}

Table~\ref{tab:calzetti_2} summarises a few of the published calibrations from
the references above; a more complete set of global calibrations can be found in
the recent review by Kennicutt \& Evans (2012). Local calibrations show
systematically higher values of $a_{\lambda_2,{\rm Type}}$ than global ones. At
24--25\,$\mu$m, $a_{24,{\rm local}}/a_{25,{\rm global}}\sim1.55$. This difference
cannot be attributed to the difference between \textit{L}(24) and
\textit{L}(25), which is around 2\% typically (Kennicutt {\it et al.} 2009;
Calzetti {\it et al.} 2010). The larger fraction of 24\,$\mu$m emission that
needs to be added to either FUV or H$\alpha$ in local SFR measurements may
simply reflect the fact that regions within galaxies are probing stellar
populations over shorter timescales ($\tau\sim$100\,Myr or smaller) than global
SFR measurements ($\tau>$ many~Gyr). The dust emission traces this difference
accordingly (Kennicutt {\it et al.} 2009). From Table~\ref{tab:calzetti_1}, the
ratio of the calibration constants for SFR(TIR) at 100\,Myr and 10\,Gyr is 1.75,
close to the observed value of 1.55. Differences in the mean dust temperature, that
is likely to boost the \textit{L}(24) in sub-galactic regions, may account for
the remaining discrepancy.

\subsubsection{Indicators based on other processes}
SFR indicators based on non-thermal (synchrotron) radio and 
X-ray emission represent more indirect ways of probing star formation in 
galaxies. 

In the case of synchrotron emission, the basic mechanism is the production and
acceleration of cosmic rays in supernova explosions; since the supernova rate is
directly related to the SFR, we should be able to use the synchrotron luminosity
as a proxy for the SFR. There is, however, an added complication in that the
non-thermal luminosity  depends not only on the mean cosmic ray production per
supernova, but also on the galaxy's magnetic field (e.g., Rybicki \& Lightman
2004). The case for SFR(sync) is helped by the well-known IR-radio correlation
(e.g., Yun {\it et al.} 2001): if the IR is correlated with both SFR and radio
emission, then SFR and radio emission are correlated among themselves. SFR(sync)
calibrations can only be derived empirically (Condon 1992; Schmitt {\it et al.}
2006; Murphy {\it et al.} 2011), because of the complexity of the relation
between the SFR and the underlying physical mechanism; a recent derivation
consistent with our IMF can be found in Murphy (2011). 

A similarly indirect relation exists between SFR and X-ray luminosity. In
star-forming galaxies, the X-ray luminosity is produced by high-mass X-ray
binaries, massive stars, and supernovae, but non-negligible contributions from
low-mass X-ray binaries are also present. The latter are not directly related
to recent star formation, and represent a source of uncertainty in the
calibration of SFR(X-ray). Because of the difficulty of establishing the
frequency and intrinsic luminosity of each X-ray source  (related or unrelated
to current star formation) from first principles, the SFR(X-ray) calibrations
have been derived empirically, and examples are given in Ranalli {\it et al.}
(2003), Persic \& Raphaeli (2007), and Mineo {\it et al.} (2012). Care should be
taken when comparing these published calibrations, obtained for a Salpeter IMF,
with those reported in this summary, which are based on a Kroupa IMF. 

\subsection{A word about the stellar initial mass function and \newline stochastic sampling}
\label{IMF}

All calibrations listed in this section make the implicit assumption that the
stellar IMF is constant across all environments and given by the double-power
law expression (Kroupa 2001):
{\begin{eqnarray}
\label{eq:calzetti_12}
\chi(M) = dN/dM &=& A\ M^{-1.3}~~~~~~~~~~(0.1 \le M(M_{\odot}) \le 0.5)\nonumber \\
                &=& 0.5\ A\ M^{-2.3}~~~~~(0.5 \le M(M_{\odot}) \le 100)
\end{eqnarray}
where $\chi(M)$ is the number of stars between $M$ and $M+dM$. The stellar mass
distribution and total stellar mass produced by this expression are not
significantly different from that produced by the log-normal expression
proposed by Chabrier (2003). The Kroupa IMF expression produces a smaller number
of low-mass stars than the Salpeter (1955) IMF, which has been customarily
represented with a single power law with slope $-2.35$ between 0.1 and
100\,$M_{\odot}$. Since the majority of SFR indicators trace massive stars, a
calibration based on the Kroupa IMF can be converted to one using the Salpeter
IMF simply by multiplying the calibration constant by 1.6. 

The assumption that the IMF is constant and universal is justified by many
observational results, although these are generally rather uncertain, especially
at the high-mass end (review by Bastian {\it et al.} 2010). There is still the
possibility of variation in some extreme (in terms of density, SFR, or other)
environments, and arguments both in favour of and against variations have been
brought forward by many different authors. To gauge the impact of a different
IMF assumption on our SFR calibrations, we can adopt a modified Kroupa IMF, with
the maximum stellar mass set to 30\,$M_{\odot}$, instead of 100\,$M_{\odot}$. The
new calibration constants, for selected timescales, are listed in
Table~\ref{tab:calzetti_1}. The constants change by factors 1.4, 1.5, and 5.6
for SFR(UV), SFR(TIR), and SFR(H$\alpha$), respectively. The change for the
H$\alpha$ calibration is the largest of all; it is larger than the UV one by a
factor of four, simply because significant UV emission is produced by stars down
to $\sim$5\,$M_{\odot}$, but significant ionising photon flux is produced only by
stars more massive than $\sim$20\,$M_{\odot}$. In addition, it takes slightly
longer (10\,Myr for the upper mass limit of 30\,$M_{\odot}$versus 6\,Myr for
100\,$M_{\odot}$) for the ionising photons to reach their asymptotic value. The
changes for the UV and TIR calibration constants are similar to each other. 

In contrast with the results just discussed, the {\em mean} stellar mass for the
Kroupa IMF is $\langle M\rangle$$\sim$0.6\,$M_{\odot}$, with less than 10\%
difference between using 100\,$M_{\odot}$ or 30\,$M_{\odot}$ as stellar upper mass
limit. This makes tracers based on the mean stellar mass of a system
(Equation~\ref{eq:calzetti_1}) more robust than those based on tracing the most
massive stars.

Even if the IMF is universal, individual systems may show departures from this
condition, based on simple arguments of sampling.

If we consider a single-age, very young stellar cluster, we can ask what the
minimum mass is that this cluster needs to have so that at least one star with
mass 100\,$M_{\odot}$ is formed. The mass is 2.8$\times$10$^5$\,$M_{\odot}$, which
is a large value, only achieved by some of the most massive star clusters known.
As a comparison, if the maximum stellar mass is 30\,$M_{\odot}$, full sampling of
the IMF, meaning that at least one 30\,$M_{\odot}$ star is formed, is achieved
with a cluster mass of 1.7$\times$10$^4$\,$M_{\odot}$.

Under these circumstances, it is not uncommon that studies that involve low
SFRs, either because the region considered is small and/or inefficient at
forming stars, or because the galaxy has a low overall SFR, are subject to the
effects of stochastic sampling, i.e., the stellar IMF is randomly, not fully,
sampled. The impact of stochastic sampling is higher for the most massive stars,
since there are proportionally less massive stars than low-mass ones. From the
Kroupa IMF expression above, only 11\% of all stars, by number, have masses
above 1\,$M_{\odot}$, although these stars represent 56\% of the total mass. 

Stochastic sampling has a larger impact on tracers of ionising photons than on
tracers of UV continuum light, for the same reason that a low upper limit in
stellar mass has. Clear evidence for this is shown by the so-called extended UV
(XUV) disks of galaxies, as revealed by \textit{GALEX}. The original hypothesis
that these XUV disks, bright in the UV but faint in H$\alpha$, could be due to
peculiar IMFs (e.g., deficient in high-mass stars) has been replaced by the
finding that the IMF is stochastically sampled in these low-SFR areas (Goddard
{\it et al} 2010; Koda {\it et al.} 2012). The models of Cervi\~no {\it et al.}
(2002), renormalised to the Kroupa IMF (Equation~\ref{eq:calzetti_12}), show
that a star cluster with mass $\sim$1$\times$10$^4$\,$M_{\odot}$ will be subject
to sufficient stochastic sampling that a scatter as large as 20\% can be
expected in the measured ionising photon flux. The scatter increases
dramatically for decreasing cluster mass, and becomes as large as 70\% for a
cluster mass $\sim$1$\times$10$^3$\,$M_{\odot}$.  This poses a practical
limitation of SFR$\gtrsim$0.001\,$M_{\odot}$~yr$^{-1}$ for the use of SFR
indicators based on the ionising photon flux if a 20\% or less uncertainty is
desired; a similar uncertainty value for the UV is obtained at
SFR$\gtrsim$0.0003\,$M_{\odot}$~yr$^{-1}$, or about 3.5 times lower than when
using tracers of ionising photons (Lee {\it et al.} 2009, 2011). 

%
%

\section{The nature of `diffuse' light in galaxies}
\label{DIFFUSE}

In the previous sections, I have attempted to discriminate SFR calibrations
applicable to whole galaxies from those applicable to regions within galaxies.
As already mentioned at the beginning of this chapter, whole galaxies are, in
first approximation, isolated systems. As long as they are calibrated and used
in a self-consistent manner, most SFR indicators should yield similar values,
and reflect the actual rate of recent star formation in a galaxy. 

Regions within galaxies, instead, are emphatically not isolated. Most young star
clusters disperse quickly (infant mortality due to gas expulsion), and they
continue to disperse as they evolve, due to both stellar evolution and a variety
of dynamical effects that include two-body relaxation, tidal stripping,
large-scale shocks, etc. (Lamers {\it et al.} 2010). Models agree that  as many
as 80\%--90\% of clusters dissolve within the first 10--20\,Myr of life, and
their stars become part of the diffuse field, although the same models tend to
disagree on the level of evolution at later stages  (Fall {\it et al.} 2005;
Lamers {\it et al.} 2005; Chandar {\it et al.} 2010). Cluster evolution and
dispersal within the first few tens of Myr is what this review is mostly
concerned with, since this can influence the derivation and interpretation of
local SFRs. 

Analysis of the \textit{HST} UV spectra of star clusters and of the intracluster
diffuse stellar light in the starburst regions of nearby galaxies has revealed
marked differences between the two stellar populations. Star clusters show clear
signatures of the presence of O star wind features, signalling the existence of
stars more massive than $\sim$30\,$M_{\odot}$ in the clusters. Conversely, the
intracluster population systematically lacks those features (Tremonti {\it et
al.} 2001; Chandar {\it et al.} 2005). Sufficient area is covered in each galaxy
that stochastic sampling in the diffuse light regions is not an issue. The
significant difference in the spectral features of cluster and intracluster
spectra excludes the possibility that the intracluster UV light is scattered
light which originates from the clusters themselves.  Only two scenarios, thus,
appear in agreement with the data: (1) stars form locally in the diffuse field,
but either they have a different IMF than that of the clusters or they form in
small clusters that systematically lack massive stars (Meurer 1995; Weidner {\it
et al.} 2010); (2) most stars form in clusters, and the clusters dissolve over
7--10\,Myr (Tremonti {\it et al.} 2001; Chandar {\it et al.} 2005). 

Option (2) may be the most viable of the two scenarios, in light of what has
been discussed earlier in this section. An important consideration is that
option (1), also referred to as in-situ star formation, necessarily implies
either a different IMF or a different cluster mass function between the clusters
and the field. 

\textit{GALEX} has surveyed many local star-forming galaxies in the UV, at
0.153\,$\mu$m (FUV) and 0.231\,$\mu$m (NUV), across their entire disk. One of the
most striking results is that  the FUV$-$NUV colours of the arm regions are in
general significantly bluer than those of the interarm regions (D. Thilker,
private communication), despite the arms usually containing more dust than the
interarms, roughly 1\,mag more in the $I$-band (White {\it et al.} 2000; Holwerda
{\it et al.} 2005). These red interarm colours, if not attributable to dust
reddening, can result from evolving stellar populations that are diffusing from
the arm regions (e.g., Pellerin {\it et al.} 2007). For example, in the galaxy
NGC\,300, located only 2\,Mpc away, the UV colours are consistent with a dominant
interarm population that is devoid of stars younger than 10--20\,Myr. Another
scenario for the UV light in the interarm regions of disk galaxies is
dust-scattered light diffusing from the spiral arms (Popescu {\it et al.} 2005),
although the significantly redder colours in the interarm regions could pose a
challenge to this interpretation. Unambiguous evidence for scattered UV photons
by dust has been found in the starburst-driven outflows of the two nearby
galaxies M\,82 and NGC\,253 (Hoopes {\it et al.} 2005).

The UV light at 0.16\,$\mu$m of a stellar population undergoing constant star
formation for the past 100\,Myr is contributed for $\sim$70\% and $\sim$30\% by
stars younger and older than 10\,Myr, respectively. We assume for simplicity that
all stars younger than 10\,Myr populate spiral arms, and those older than that
are located in the interarm regions. Adopting extinction values of $A_I$=1.5\,mag
and 0.5\,mag in the spiral and interarm regions, respectively, as determined by
Holwerda {\it et al.} (2005), and a very simple dust geometry, the {\it
observed} 0.16\,$\mu$m UV emission becomes 40\% and 60\% contributed by stars
younger and older than 10\,Myr, respectively, almost reverting the intrinsic
ratios. The observed ratio is about 50\%--50\% in NGC\,5194 and NGC\,3521 (Liu
{\it et al.} 2011). Dust geometry plays a crucial role in this case, and almost
about any value of the observed UV fraction from the two components can be
obtained by varying the dust geometry, within a reasonable range for the
geometrical distributions of dust and stars (see section~\ref{DUSTATTEN}). 

Whether due to dust scattering and/or ageing stellar populations diffusing from
the spiral arms and/or some in-situ star formation, or a combination of all
three, the UV light in the inter-arm regions of spiral galaxies displays a more
complex nature than that of the arms. This consideration suggests caution in
using standard calibrations of the SFR(UV) in spatially-resolved studies of
disks, especially when the targeted regions include inter-arm areas that have
not been independently confirmed to be actively star-forming. 

The escape of ionising photons from H{\sc ii} regions has been already discussed
in Section~\ref{SFRION}. It is at the level of 40\%--60\% for the {\it observed}
H$\alpha$ emission (e.g., Ferguson {\it  al.} 1996; Thilker {\it et al.} 2002;
Oey {\it et al.} 2007), but gets reduced to $\sim$30\% when differential
extinction in the H$\alpha$ within and outside the H{\sc ii} regions is
accounted for (Crocker {\it et al.} 2012). The nature of the diffuse H$\alpha$
in galaxies has been the subject of extensive studies by a number of authors,
with still some open questions (e.g., Witt {\it et al.} 2010). For the purpose
of measuring SFRs, we need to consider two effects. Firstly, the leakage of
ionising photons from star-forming regions will impact SFR($Q(H^o$)), reducing
it roughly by 30\% (trends with luminosity, gas density, etc, are only now
starting to be investigated, see Pellegrini {\it et al.} 2012). Secondly, weak
recombination line emission will appear in regions that are not star-forming; in
sensitive surveys, this emission could be mistaken for faint in-situ star
formation. This second effect should not be underestimated, as ionising photons
have been shown to travel as far as about 1\,kpc from their point of origin. 

The non-discriminating nature of dust in regard to the sources of heating poses
another challenge for local SFR measurements, if the IR is used as an indicator,
either alone or in combination with a UV/optical one. The UV/optical photons
produced by the stellar population of the diffuse field are generally sufficient
to heat the dust in a galaxy (Draine {\it et al.} 2007). As already mentioned in
Section~\ref{SFRIR}, the 8\,$\mu$m emission from a galaxy may be a better tracer
of B stars than recent star formation (Boselli {\it et al.} 2004; Peeters {\it
et al.} 2004), and about 30\% of the 8\,$\mu$m emission from the nearby galaxy
NGC\,628 is unrelated to star formation more recent than 100\,Myr (Crocker {\it
et al.} 2012). A similar fraction, $\sim$30\%, is recovered at 24\,$\mu$m, when
comparing the local with the global SFR indicators (see Section~\ref{SFRMIX}).
Direct attempts to quantify the fraction of 24\,$\mu$m emission unrelated to
recent star formation provide a minimum conservative value of about 20\% (Leroy
{\it et al.} 2012), with as much as 50\% of the emission heated by populations
older than $\sim$10\,Myr (Liu {\it et al.} 2011). Intermediate (0.2--2\,Gyr) age
stars can be bright in the 20--45\,$\mu$m region, significantly contributing to
the emission in this wavelength region (Verley {\it et al.} 2009; Kelson \&
Holden 2010). 

In a study of the Triangulum Galaxy (M\,33), Boquien {\it et al.} (2011)
identify a threshold of SFR/area=10$^{-2.75}$\,$M_{\odot}$~yr$^{-1}$~kpc$^{-2}$,
above which the IR emission reliably traces the current SFR; in M\,33, this
threshold delineates the spiral arms and the central $\sim$1.2\,kpc region, and
excludes most of the interarm regions, which are dominated by evolved star
heating. Dust can also be heated by UV photons leaking out of the arms into the
interarm regions (Popescu {\it et al.} 2005; Law {\it et al.} 2011). In this
case, the heating is not produced by in-situ star formation, but by photons that
have originated a large distance away. In summary, care should be taken when
attempting to measure SFRs in faint galaxy regions using the IR emission, since
this emission may be dominated by heating by old stars and/or UV photons leaking
out of H{\sc ii} regions a large distance away, instead of tracing in-situ star
formation. 

The general conclusion to be taken away from this section is that SFR
measurements {\it at any wavelength} could be tricky in regions that are not
obviously dominated by recent star formation, such as the interarm regions of
galaxies. 

%
%

\section{Dust attenuation of the stellar light}
\label{DUSTATTEN}

\subsection{General properties}
In this section, I will briefly discuss the effect of dust attenuation on both
the stellar continuum and the ionised gas line emission. Because of the
complexity of the topic, this section will be incomplete, and the interested
reader is referred to the included list of references, and to the review of
Calzetti (2001).

I make here the explicit distinction between `attenuation' (minted for such use
by Gerhardt Meurer, to my best recollection) and `extinction'. 

Extinction refers to the combined absorption and scattering (out of the line of
sight) of light by dust. The light is provided by a background point source
(star, quasar), and the dust is entirely foreground to the source. Because the
source is a background point, the distribution of the foreground dust is
irrelevant to the total extinction value (left panel of Fig.~\ref{fig03}).

Attenuation refers to the net effect of dust in a complex geometrical
distribution, where the light sources are distributed within the dust at a range
of depths, including in front of and behind it, and the dust itself can be
clumpy, smooth, or anything in between. Because both the light sources and the
dust have extended distributions, their relative location has a major impact on
the net absorbed and scattered light, the latter now including scattering {\it
into}, as well as out of, the line of sight (right panel of Fig.~\ref{fig03}).
Dust scattering into the line of sight has the effect of producing a greyer
overall attenuation than if only scattering out of the line of sight were
present, and the emerging SED will be bluer. This is the typical situation
encountered when studying galaxies or extended regions within galaxies.

 \begin{figure}  
\includegraphics[width=\textwidth]{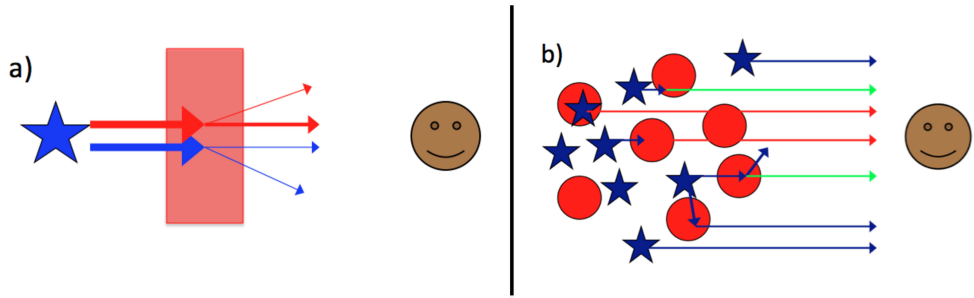}
\caption{The left (a) and right (b) panels are cartoon representations of: a) a
point source (star) behind a screen of dust; and b) an extended distribution of
stars mixed with clumps of dust. The left panel shows the typical configuration
that enables measures of {\it extinction curves}: a single illuminating source
located in the background of the dust. The dust extinguishes the light via
direct absorption and via scattering out of the line of sight. The right panel
is more representative of the situation encountered whenever a complex
distribution of stars and dust is present, as found in external galaxies and
large regions. In this case, different stars may encounter different numbers of
dust clouds (differences in optical depth), and some stars can be entirely
embedded in dust (internal extinction) or be completely foreground to the dust
distribution. Scattering of the light by dust both {\it into} and {\it out of}
the line of sight is present. Because of the more complex geometrical relation
between dust and the illuminating sources,  the net effect of dust  on the
stellar population's SED is termed {\it attenuation}.}
\label{fig03}
\end{figure}

The radiative transfer of light through dust is described by an
integro-differential equation. At UV/optical/near-IR wavelengths, the
radiative transfer equation is: 
\begin{equation} 
\label{eq:calzetti_13}
\frac{d I_{\nu}}{d \tau} = -I_{\nu} + \frac{a_{\nu}}{\pi} \int I_{\nu} \phi (\nu, {\rm cos}\ \Theta) d \Omega
\end{equation}
where $I_{\nu}$ is the light intensity, $\tau$ is the optical depth through the
dust, $a_{\nu}$ is the dust albedo (i.e., the ratio of the scattering
coefficient to the sum of the scattering and absorption coefficients), $\phi
(\nu, {\rm cos}\ \Theta) $ is the scattering phase function,and $\Theta$ is the angle
between the scattered photon and the line of sight. Expressions for both
$a_{\nu}$ and $\phi (\nu, {\rm cos}\ \Theta)$ are given in Draine (2003b). In the
equation above, we have neglected the source function, i.e., the dust emission,
which usually has small values for wavelengths shorter than a few~$\mu$m. The
first term to the right-hand side of the equation describes the decrease in
intensity of the original beam due to passage inside the dust, and the second
term is the light added to the beam by scattering into the line of sight. 

General solutions to the problem of how to remove the effects of dust from
extended systems are not available. One of the first papers to address this
issue specifically for galaxies is due to Witt {\it et al.} (1992). Since then,
many codes have been made available to the community to treat the radiative
transfer of the light produced by a stellar population through dust, with the
goal of simulating realistic SEDs of galaxies. As many such codes exist, it is
impossible to do justice to them all, and I shall avoid injustice by citing
none. 

For the simple case in which there is a single point-like light source behind a
screen of dust, Equation~\ref{eq:calzetti_13} reduces to
\begin{equation}
\label{eq:calzetti_14}
\frac{d I_{\nu}}{d \tau} = -I_{\nu} 
\end{equation}
with the well-known solution for the extinction of a stellar spectrum by 
foreground dust
\begin{equation}
\label{eq:calzetti_15}
I_{\nu}= I_{\nu}^o\  {\rm e}^{-\tau}
\end{equation}
where $I_{\nu}^o$ is the incident light and
\begin{equation}
\label{eq:calzetti_16}
\tau_{\lambda} = 0.921\ A(\lambda) = 0.921\ E(B-V)\ \kappa(\lambda)
\end{equation}
is the optical depth, which is related to the extinction curve $\kappa(\lambda)$
through the colour excess $E(B-V)$. The colour excess is a measure of the
thickness of the dust layer, while the extinction curve provides a measure of
the overall cross-section of dust to light as a function of wavelength.
Observational measures of extinction curves have been obtained only for the
Milky Way, the Magellanic Clouds, and M\,31 (Cardelli {\it et al.} 1989; Bianchi
{\it et al.} 1996; Fitzpatrick 1999; Gordon {\it et al.} 2003, 2009), because
these are the only galaxies for which individual stars can be isolated and the
extinction properties of the dust in front of them determined. For more distant
systems, `extinction' measures are more properly `attenuation' measures. 

A few other `almost' exact solutions are available for
Equation~\ref{eq:calzetti_13}, and all of them require replacing the integral on
the right-hand side of the equation with some mean or central value, so the
equation changes to a pure differential one. Mathis (1972) and Natta \& Panagia
(1984) provide an expression for the case of internal extinction, which
geometrically corresponds to a homogeneous mixture of stars and dust:
\begin{equation}
\label{eq:calzetti_17}
I_{\nu}= I_{\nu}^o\  \frac{1 - {\rm e}^{-\tau'}}{\tau'}
\end{equation}
where $\tau'$ is an effective optical depth that needs to include the mean
effects of scattering into the line of sight (Mathis 1983). 

The two cases of foreground dust (screen; Equation~\ref{eq:calzetti_15}) and
internal dust (homogeneous mixture; Equation~\ref{eq:calzetti_17}) are shown in
Fig.~\ref{fig04}. This cartoon representation uses the same input stellar SED
and dust characteristics, including the dust thickness, as described by the
colour excess $E(B-V)=0.5$, and the dust extinction curve, which I take to be
the standard Milky Way curve with $R_V=A_V/E(B-V)=3.1$ (e.g., Fitzpatrick 1999).
Despite all similarities, the different geometrical relation between stars and
dust produces dramatic differences in the output SED, as shown by the two plots
to the right hand side of Fig.~\ref{fig04}. In general, a foreground screen
produces the largest reddening and dimming of all possible dust geometrical
configurations. This is a possible choice if the goal is to maximise the impact
of dust on a stellar (or other source) SED.

A homogeneous mixture of dust and stars, conversely, produces an almost grey
attenuation, with the output SED remaining basically blue even at UV
wavelengths. Adding dust to this configuration does not change the shape of the
SED in any major way, but mostly dims it. Thus, at UV/optical wavelengths, the
SED of a mixed dust/star system will appear very similar to the SED of a dimmer,
almost dust-free system. Only a FIR measurement will be able to discriminate
among the two systems. 

\begin{figure}  
\includegraphics[width=\textwidth]{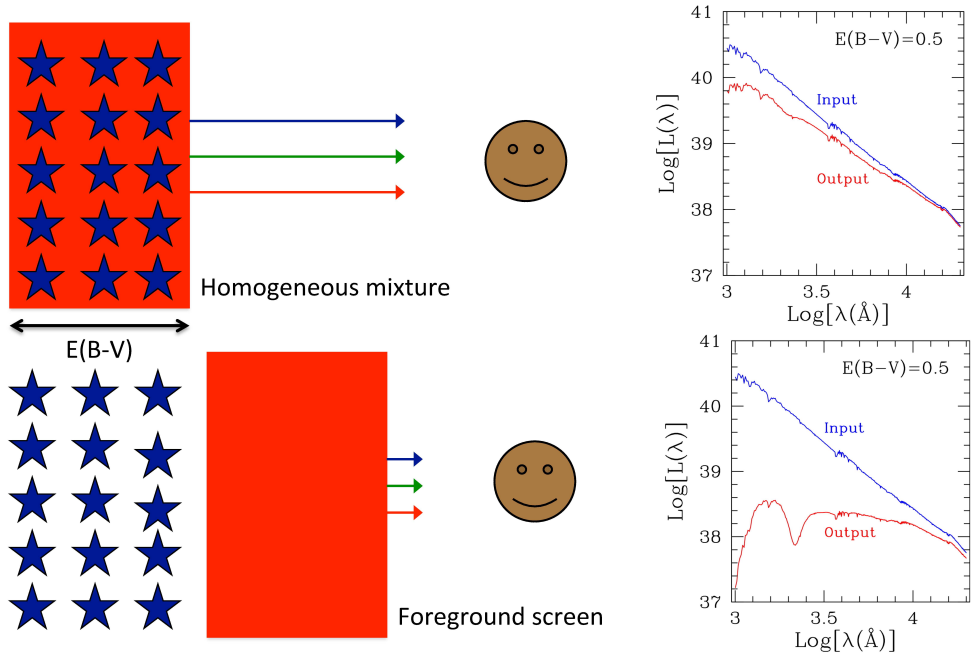}
\caption{The top and bottom panels show cartoon representations of the same
extended distribution of stars and dust, but with a different geometrical relation
between each other. In the top panel the dust and stars are homogeneously mixed,
while in the bottom panel the dust is completely foreground to the stars. The
characteristics of the stars are the same in the two panels. I have assumed that in
both cases the dust obeys the Milky Way extinction curve (which has a prominent
absorption feature at 2175\,\AA) with a thickness of $E(B-V)=0.5$. The panels to
the right show the input stellar SED, which is the same for the two cases (blue;
top spectrum), and the output SED (red;  bottom spectrum). All other
characteristics being equal, the different geometric relation between dust and
stars has considerable impact on the emerging spectrum (`Output').}
\label{fig04}
\end{figure}

The degeneracy described above is just one of the many degeneracies that are
possible when limited information is available on a galaxy SED. A notorious one
is the age/dust degeneracy, for which a young, dusty stellar population can have
a UV/optical/near--IR SED not dissimilar from that of an old, dust-free
population. Breaking of degeneracies usually requires collecting as much
information as possible about a system, including, but not limited to, emission
line luminosities, the magnitude of the 0.4\,$\mu$m break ($D_{\rm n}$(4000), 
e.g., Kauffmann {\it et al.} 2003), and the IR dust luminosity. 

A concrete example of how to constrain the dust distribution in a complex system
involves the use of hydrogen recombination lines. The intrinsic line-intensity
ratio between these lines is set by quantum mechanics, with relatively small
variations as a function of electron temperature and density if the lines are at
the wavelength of Br$\gamma$ or bluer. Measurements of at least three emission
lines, widely spaced in wavelength, probe different dust optical depths, which,
when combined, provide strong constrains on the dust geometry, at least up to
the longest wavelength probed (e.g., Calzetti 2001).  As an example,  the three
recombination lines  H$\beta$ at 0.4861\,$\mu$m, P$\beta$ at 1.282\,$\mu$m, and
Br$\gamma$ probe a factor of ten total difference in optical depth between the
bluest and the reddest line, with a factor of four between H$\beta$ and
P$\beta$. A common approach when only two lines are available, which most
typically are H$\beta$ and H$\alpha$, is to adopt a foreground dust screen, and
derive a colour excess by taking the ratio of the observed lines to that of the
intrinsic line luminosity as:
\begin{equation}
\label{eq:calzetti_18}
R_{\alpha\beta}=\frac{[L({\rm H}\alpha)/L({\rm H}\beta)]_{\rm obs}}{[L({\rm H}\alpha)/L({\rm H}\beta)]_{\rm int}} = 10^{0.4\ [\kappa({\rm H}\beta) - \kappa({\rm H}\alpha)]\ E(B-V)}
\end{equation}
where $\kappa({\rm H}\beta)$ and $\kappa({\rm H}\alpha)$ are the values of the
extinction curve evaluated at the wavelength of H$\beta$ and H$\alpha$
($\kappa({\rm H}\beta) - \kappa({\rm H}\alpha)$=1.163 for the Milky Way
extinction curve). Although this approach is rather simplistic, it appears to
work reasonably well for local star-forming galaxies, which tend to have modest
attenuation values, $A_V\sim$1\,mag, (Kennicutt 1983; Kennicutt et al. 2009), and
for `UV-bright starbursts' (see definition below;  Calzetti et al. 1996;
Calzetti 2001). 

\subsection{Application to galaxies}

Despite galaxies being difficult to treat in a general fashion, a class of
low-redshift galaxies show relatively regular behaviour in their SEDs for
increasing dust content. I term these galaxies `UV-bright starbursts', where we
use `UV-bright' to discriminate them from LIRGs and ULIRGs: the latter are
characterised by a centrally concentrated region of star formation
occupying\newpage\noindent the inner few hundred parsecs, with 90\% or more of
their energy output emerging in the IR. We use the term `starbursts' to
discriminate them from `normal star-forming' galaxies, these being characterised
by widespread star formation across the disk with a relatively low SFR surface
density, i.e., such that SFR/area $<$ 0.3--1\,$M_{\odot}$~yr$^{-1}$\,kpc$^{-2}$.
UV-bright starbursts in the local Universe are galaxies in which the central
(inner $\approx$1-2\,kpc) starburst dominates the light output at most
wavelengths, but which are still sufficiently transparent that a significant
fraction of their energy emerges in the UV. 

The UV spectral slope, $\beta$, measured in the range $\sim$0.13--0.26\,$\mu$m,
of local UV-bright starbursts is correlated with the colour excess $E(B-V)$, in
the sense that higher values of the colour excess produce redder UV SEDs for
these galaxies (Calzetti {\it et al.} 1994). The UV spectral slope of these
galaxies is also correlated with the infrared excess, measured by the ratio
\textit{L}(TIR)/\textit{L}(UV) (Meurer {\it et al.} 1999); this correlation was
termed the IRX-$\beta$ relation by the original authors, where IRX stands for
`infrared excess'. In recent years, with the wealth of UV imaging data on local
galaxies produced by \textit{GALEX}, it has become customary to replace the UV
spectral slope $\beta$ with the UV colour FUV$-$NUV, but the sense of the
correlations has remained unchanged. The interpretation of both correlations is
straightforward: larger amounts of dust, as traced by the colour excess
$E(B-V)$, produce both larger reddening, traced by $\beta$, and larger total
attenuations, traced by \textit{L}(TIR)/\textit{L}(UV), in the starbursts' SEDs.
The power of such simple correlations, especially the IRX-$\beta$ one, can be
immediately appreciated: at high redshift, rest-frame UV spectral slopes are
more immediately measurable than total attenuations, since they only require the
acquisition of an (observer-frame) optical/near-IR spectrum or colour. The
IRX-$\beta$ correlation is, indeed, obeyed by high-redshift starburst galaxies
as well (Reddy {\it et al.} 2010, 2012). 

Two other important characteristics of the dust attenuation trends in local
UV-bright starbursts are: (1) the absence of the 0.2175\,$\mu$m `bump' (a common
feature in the Milky Way extinction curve), which may be an effect of
destruction of the carriers; and (2) the fact that the ionised gas emission
suffers about twice the attenuation of the stellar continuum (Calzetti {\it et
al.} 1994). This second characteristic appears to be present also in starburst
galaxies at high redshift (e.g., Wuyts {\it et al.} 2011). 

In terms of dust attenuation, local UV-bright starbursts behave as if the dust
were located in a clumpy shell surrounding the starburst region (Calzetti {\it
et al.} 1994; Gordon {\it et al.} 1997; Calzetti 2001). With this simple
geometry, dust can be treated as a foreground screen, and the attenuation
described as (Calzetti {\it et al.} 2000):
\begin{equation}
\label{eq:calzetti_19}
f_{\rm int}(\lambda) = f_{\rm obs}(\lambda) 10^{0.4\ \kappa^{\rm e}(\lambda)\ E(B-V)_{\rm star}}
\end{equation}
where $\kappa^{\rm e}(\lambda)$ is an effective attenuation curve to be applied 
to the observed stellar continuum SED $f_{\rm obs}(\lambda)$ of a starburst 
galaxy to recover the intrinsic SED $f_{\rm int}(\lambda)$, and with expression:
\begin{eqnarray}
\label{eq:calzetti_20}
\kappa^{\rm e}(\lambda)& =& 2.659\ (-1.857 + 1.040/\lambda) + 4.05~~~~(0.63\,\mu{\rm m} \le \lambda \le 2.20\,\mu{\rm m}) \nonumber \\
         &=& 2.659\ (-2.156 + 1.509/\lambda - 0.198/\lambda^2 + 0.011/\lambda^3) + 4.05 \nonumber \\
         &  & ~~~~~~~~~~~~~~~~~~~~~~~~~~~~~~~~~~~~~~~(0.12\,\mu{\rm m} \le \lambda < 0.63\,\mu{\rm m}) 
\end{eqnarray}
and $E(B-V)_{\rm star}$ is the stellar continuum colour excess, which is smaller 
than that of the ionised gas, as follows:
\begin{equation}
\label{eq:calzetti_21}
E(B-V)_{\rm star}=0.44 \  E(B-V)_{\rm gas}
\end{equation}
$E(B-V)_{\rm gas}$ is the same as the $E(B-V)$ in Equation~\ref{eq:calzetti_18}. 

In the same spirit, the IRX-$\beta$ relation has been given as (Meurer {\it et
al.} 1999; Calzetti 2001): 
\begin{equation}
\label{eq:calzetti_22}
{\rm log_{10}} \biggl[ \frac{1}{1.68} \biggl(\frac{L({\rm TIR})}{L({\rm UV})}\biggr) + 1 \biggr] = 0.84 (\beta - \beta_{\rm o})
\end{equation}
where \textit{L}(UV) is centred around 0.15--0.16\,$\mu$m, and $\beta_{\rm o}$ is the
intrinsic (unattenuated) UV slope of the galaxies, with typical values
$\beta_{\rm o}\sim-2.2$ to $-2.3$, for constant star formation. An example of
Equation~\ref{eq:calzetti_22} is given in Fig.~\ref{fig05}, together with the 
data originally used to derive it. 

\begin{figure}  
\includegraphics[width=0.95\linewidth]{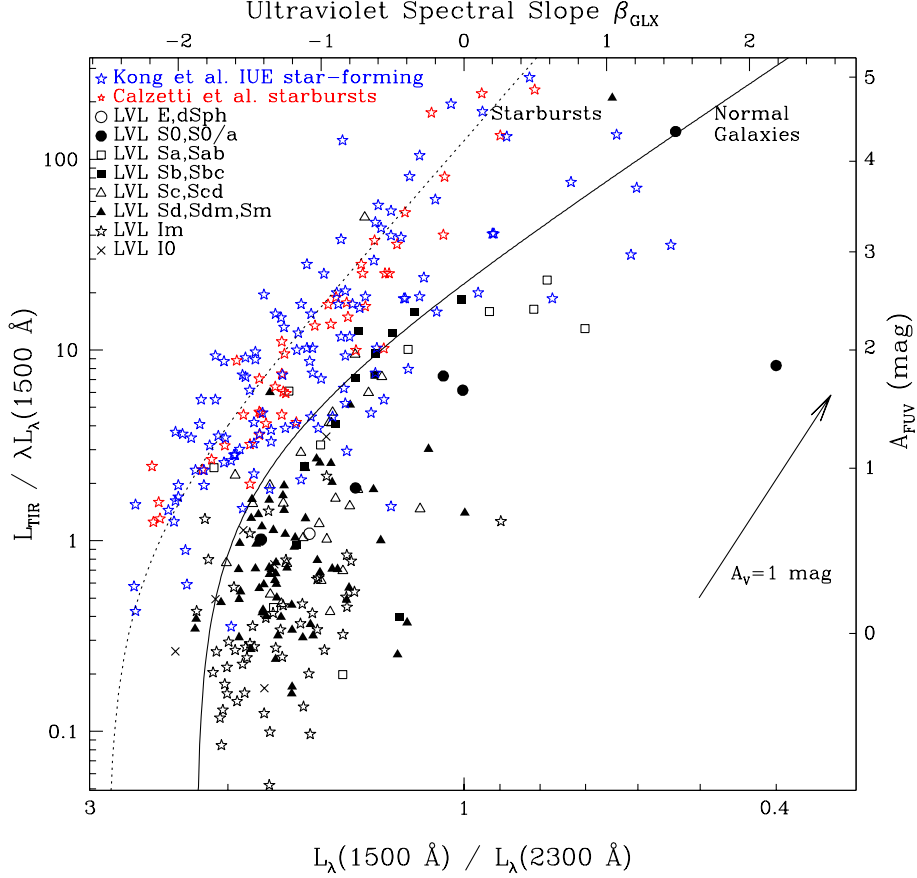}
\caption{The IRX-$\beta$ plot for local starburst and star-forming galaxies,
from Dale {\it et al.} (2009). The vertical axis is the IR excess over the UV,
where the UV is the \textit{GALEX} FUV (0.15\,$\mu$m) band. The horizontal axis
is the \textit{GALEX} FUV$-$NUV colour, expressed as luminosity ratio, with the
corresponding values of the UV spectral slope $\beta$ shown at the top of the
plot. The red points are the UV-bright starburst galaxies used by Meurer {\it et
al.} (1999) to derive the IRX-$\beta$ relation, shown by the dotted line
(Equation~\ref{eq:calzetti_22}). The blue and black points give the location of
normal star-forming galaxies from samples of the local Universe. These galaxies
have a much larger spread in the IRX-$\beta$ plane than the UV-bright
starbursts, and typically lower IR excesses at constant UV slope. Their mean
trend is shown by the continuous line. An A$_V$=1\,mag attenuation vector is also
shown. Reproduced with permission from Dale {\it et al.} (2009).}
\label{fig05}
\end{figure}

Deviations from the simple foreground geometry that can be used for UV-bright
starbursts were noted as soon as additional classes of galaxies started to be
investigated for systematic trends with dust attenuation. Local LIRGS and
ULIRGs, for instance, mostly fall above the locus defined by
Equation~\ref{eq:calzetti_22} in the IRX versus $\beta$ plot (e.g., Goldader
{\it et al.} 2002): typically these galaxies have large IR excesses for their UV
slopes. The same trend is observed in high-redshift ULIRGs (Reddy {\it et al.}
2010). Geometries that can account for this behaviour include shells of
scattering dust and clumps (Calzetti 2001), which can be realised if the dust is
located in close proximity to the heating sources, as would be the case in the
high-density central regions of the IR-luminous galaxies. 

Normal star-forming galaxies, as defined above, also deviate from the locus
defined by Equation~\ref{eq:calzetti_22}, as do star-forming regions within
these galaxies. They generally tend to be located below the starburst curve,
i.e., to have low IR excesses for their UV colours, and to have a large spread,
about a factor 5--10 larger than the UV-bright starbursts (Fig.~\ref{fig05}).
This trend has been reported by a large number of authors who have analysed
samples of local galaxies and regions within galaxies (Buat {\it et al.}  2002,
2005; Gordon {\it et al.}  2004; Kong {\it et al.}  2004; Calzetti {\it et al.}
2005; Seibert {\it et al.}  2005; Boissier {\it et al.}  2007; Dale {\it et al.}
2009; Boquien {\it et al.}  2009, 2012). Similar deviations have been reported
also for some high-redshift galaxies (Reddy {\it et al.} 2012). The
interpretation for both the shift towards lower IR excesses and the larger
spread than starbursts varies from author to author, and includes: (1) a range
of ages in the dominant UV populations; (2) scatter and variations in the dust
geometry and composition; or (3) a combination of both. Perhaps, the third
interpretation may ultimately be the correct one. Unlike starbursts, in which a
more or less causally connected region dominates the energy output, normal
star-forming galaxies are a collection of unconnected star-forming regions, each
with its own dust geometry and mean age, amid an evolving, but not necessarily
UV-faint, diffuse stellar population (Calzetti 2001). In such systems, the UV
colour, which has the smallest leverage by covering the shortest wavelength
range, will be very sensitive to influences from stellar populations and dust
geometry variations (e.g., Kong {\it et al.} 2004; Calzetti {\it et al.} 2005;
Boquien {\it et al.} 2009; Hao {\it et al.} 2011). Whether one or the other
factor predominates, and under which conditions, is still subject of
investigation, and it would be premature to provide here a definite answer. 

%
%

\section{Lessons learned}

In line with the rest of this chapter, I will summarise the lessons learned on
SFR indicators by separating the global, whole-galaxies case from the one
describing local, sub-galactic regions.

\subsection{Global SFR indicators}

As the integrated light from galaxies is a weighted average of the most luminous
contributors (i.e., star-forming regions), it is perhaps not surprising that
global SFR indicators show a high level of consistency, as summarised in
Kennicutt \& Evans (2012). As long as the assumptions for the stellar IMF are
factored in the calibration constant, stochastic sampling of the IMF is not an
issue, and both the dust-obscured and dust-unobscured star formation are
accounted for, different SFR indicators should yield similar answers. 

All other conditions being equal, SFR indicators that are sensitive only to
short timescales, i.e., only probe the presence of short-lived, massive stars,
should be preferred to long-timescale ones. Examples of short-timescale SFR
indicators are those using ionised gas tracers (e.g., H$\alpha$). Conversely,
the IR  probes emission from stellar populations covering a large range of ages,
and its use will depend on the dominant stellar population contributing to the
IR emission and on the required accuracy for the SFR measure: as shown in
Table~\ref{tab:calzetti_1}, the calibration constant changes by a factor of 1.75
if the timescale of the star formation changes from 100\,Myr (e.g., a LIRG or
more luminous galaxy) to 10\,Gyr (a normal star-forming galaxy).

At high attenuation values, which generally correspond to high SFR values, about
a few times 10\,$M_{\odot}$~yr$^{-1}$ or above, the dust starts competing with
the gas for Lyman continuum photon absorption. Combining an ionised gas tracer
(e.g., H$\alpha$) with a dust emission tracer (e.g., 24\,$\mu$m) is likely to not
only mitigate this problem, but also provide a general answer to the question
of how to correct UV/optical tracers for the effects of dust attenuation. Mixed
SFR indicators (H$\alpha+$24\,$\mu$m, UV$+$24\,$\mu$m, etc.) have, indeed, broad
applicability in all cases where stochastic IMF sampling is not a concern.

At low SFR values, below $\sim$10$^{-3}$\,$M_{\odot}$~yr$^{-1}$, stochastic
sampling of the IMF affects the use of ionised gas tracers as SFR indicators.
The longer-lived UV emission may thus become a preferable choice, as long as it
is corrected for the effects of dust attenuation. Even the UV, however, is not a
`panacea', since stochastic IMF sampling starts affecting the UV emission barely
a factor three to four below the SFR level of the ionised gas. 

Exclusive use of the UV (even after dust attenuation corrections) may complicate
the discrimination between star-forming galaxies and post-star-forming galaxies
(e.g., Weisz {\it et al.} 2012), i.e., galaxies whose active star formation
terminated many tens of Myr ago and for which the use of any of the calibration
constants in Table~\ref{tab:calzetti_1} will only yield a lower limit. 

\subsection{Local SFR indicators}

Unlike whole galaxies, regions {\it within} galaxies are not isolated systems,
and a variety of issues needs to be considered when attempting to convert any
luminosity into a SFR. Evolution and mixing of stellar populations and the
ability of stellar continuum light and ionising photons to leak out of
star-forming regions and travel to distances of 1\,kpc, and possibly more, are
important effects that need to be taken into account for deriving local SFRs.
Most of the problems do not reside with the regions of star formation proper,
but with the faint regions that may have little or no in-situ star formation.
These cumulatively termed `non-in-situ star formation' regions can still emit at
many wavelengths, including H$\alpha$, because of leakage from surrounding
areas. 

The different estimates on the importance of these effects that can be found in
the literature (e.g., more than a factor of two difference between Liu {\it et
al.} 2011 and Leroy {\it et al.} 2012) attest to the complexity of the issue. An
example of the problems induced by inaccurate estimates of local SFRs is given
in Calzetti {\it et al.} (2012). Here, it is shown that the relation between the
SFR and cold gas surface densities, also known as, the Schmidt-Kennicutt law,
strongly depends on the treatment of the `non-in-situ' star formation. 

A safe approach, at least for now, is to measure SFRs only in regions where
there is ample independent evidence that star formation is actually occurring,
such as in spiral arms and in the central regions of many galaxies. The study of
M\,33 has shown that dust heating is mainly powered by recent star formation in
these locales, down to SFR/area$\sim$0.002\,$M_{\odot}$~yr$^{-1}$~kpc$^{-2}$
(Boquien {\it et al.} 2011).  

Even when star-forming regions or areas have been identified, care should be
taken with how the different SFR indicators are applied: the one to choose for a
specific case may depend on the star formation timescale of interest. In
general, the shorter the timescale, the lower the dependence of the SFR
indicator on the evolution of the stellar population. Ionising photon tracers
may `fit the bill',  although leakage of ionising photons out of star-forming
regions will need to be accounted for. 

%
%

\section*{Acknowledgments}
I am extremely grateful to the organisers of this Winter School, Johan Knapen
and Jesus Falc\'on-Barroso, for inviting me and for the opportunity to deliver
these lectures. I am also grateful to the Instituto de Astrof\'isica de Canarias
and its Director, Prof. Francisco S\'anchez, for the hospitality. Many of the
results presented in this manuscript are the products of the SINGS
(\textit{Spitzer} Infrared Nearby Galaxies Survey), KINGFISH (Key Insights in
Nearby Galaxies: a Far Infrared Survey with \textit{Herschel}), and LVL (Local
Volume Legacy) collaborations, to whose members I am profoundly indebted. 

I also want to thank my long-time, long-distance collaborator Robert C.
Kennicutt; scientific discussions with him are always enlightening. We often end
up with friendly disagreements, and both of our healths have benefited from
never residing closer than about 2000\,km from each other. Finally, parts of
this manuscript have been improved thanks to discussions with another
long-time collaborator, John S. Gallagher. 

The preparation of this manuscript was supported in part by the NASA-ADAP grant
NNX10AD08G and in part by the NASA \textit{Herschel} grant JPL-1369560 to the 
University of Massachusetts. 

%
%

\end{document}